\documentclass[
twocolumn,
preprint,
preprintnumbers,
linenumbers,
amsmath,amssymb,
]{openjournal}

\usepackage{graphicx}% Include figure files
\usepackage{dcolumn}% Align table columns on decimal point
\usepackage{bm}% bold math
\usepackage{stfloats}
\usepackage{orcidlink}
\usepackage{amsmath}
\usepackage{amssymb}

\usepackage{siunitx} % provides a set of tools to typset quantities in a reasonable way
\usepackage{multirow}

\usepackage{booktabs}   % nice table rules
\usepackage{csvsimple}  % load CSV/TXT tables
\usepackage{tabularx} % in the preamble
\usepackage{makecell}

\newcommand{\MARZ}{\texttt{MARZ}}
\newcommand{\PypeIt}{\texttt{PypeIt}}

\definecolor{jackgreen}{rgb}{0.0, 0.5, 0.0}

\definecolor{reviewerblue}{rgb}{0.0, 0.0, 0.5}
\newcommand{\reviewer}[1]{{\color{reviewerblue}\bf[REVIEWER: #1]}}

\begin{document}

\title{
The Carousel Lens I:\\
A Spectroscopic Survey of the Carousel Lens Field
}

\author{Jackson H. O'Donnell$^{\hyperlink{inst:UCSC}{1},\hyperlink{inst:SCIPP}{2}}$\orcidlink{0000-0003-4083-1530}}
 \email{jhodonnell@ucsc.edu}
\author{Demetrius Y. Williams$^{\hyperlink{inst:UCSC}{1},\hyperlink{inst:SCIPP}{2}}$\orcidlink{0000-0003-1042-1995}}
 \email{dywilliams629@gmail.com}
\author{Tesla E. Jeltema$^{\hyperlink{inst:UCSC}{1},\hyperlink{inst:SCIPP}{2}}$\orcidlink{0000-0001-6089-0365}}
\author{William Sheu$^{\hyperlink{inst:UCLA}{3}}$\orcidlink{0000-0003-1889-0227}}
\author{Felipe Urcelay$^{\hyperlink{inst:OXF}{4}}$\orcidlink{0009-0009-0407-2419}}
\author{Xiaosheng Huang$^{\hyperlink{inst:USF}{5},\hyperlink{inst:LBNL}{6}}$\orcidlink{0000-0001-8156-0330}}
\author{Tucker Jones$^{\hyperlink{inst:UCD}{7}}$\orcidlink{0000-0001-5860-3419}}
\author{Karl Glazebrook$^{\hyperlink{inst:CASS}{8},\hyperlink{inst:ARC}{9}}$\orcidlink{0000-0002-3254-9044}}
\author{Tania  M. Barone$^{\hyperlink{inst:CASS}{8},\hyperlink{inst:ARC}{9}\hyperlink{inst:CFA}{10}}$\orcidlink{0000-0002-2784-564X}}
\author{Aleksandar Cikota$^{\hyperlink{inst:NOIR}{11}}$\orcidlink{0000-0001-7101-9831}}
\author{Fuyan Bian$^{\hyperlink{inst:ESO}{12},\hyperlink{inst:CAS}{13}}$\orcidlink{0000-0002-1620-0897}}
\author{Christopher J. Storfer$^{\hyperlink{inst:LBNL}{6},\hyperlink{inst:IFA}{14}}$\orcidlink{0000-0002-0385-0014}}
\author{Daniel J. Ballard$^{\hyperlink{inst:SYD}{15}}$\orcidlink{0009-0003-3198-7151}}
\author{Gabriel Caminha$^{\hyperlink{inst:TUM}{16},\hyperlink{inst:MPA}{17}}$\orcidlink{0000-0001-6052-3274}}
\author{Glenn G. Kacprzak$^{\hyperlink{inst:CASS}{8},\hyperlink{inst:ARC}{9}}$\orcidlink{0000-0003-1362-9302}}
\author{Themiya Nanayakkara$^{\hyperlink{inst:CASS}{8},\hyperlink{inst:ARC}{9}}$\orcidlink{0000-0003-2804-0648}}
\author{Nandini Sahu$^{\hyperlink{inst:ARC}{9},\hyperlink{inst:NSW}{18}}$\orcidlink{0000-0003-0234-6585}}
\author{Hannah Skobe$^{\hyperlink{inst:CMU}{19}}$\orcidlink{0000-0003-0516-3485}}
\author{Anowar J. Shajib$^{\hyperlink{inst:UChi}{20},\hyperlink{inst:KICP}{21},\hyperlink{inst:IUD}{22}}$\orcidlink{0000-0002-5558-888X}}
%\author{Sherry Suyu$^{\hyperlink{inst:TUM}{16},\hyperlink{inst:MPA}{17}}$\orcidlink{0000-0001-5568-6052}}
\author{Kim-Vy Tran$^{\hyperlink{inst:CFA}{10}}$\orcidlink{0000-0001-9208-2143}}
\author{Keerthi Vasan G. C.$^{\hyperlink{inst:CO}{23}}$\orcidlink{0000-0002-2645-679X}}

\affiliation{%
 \hypertarget{inst:UCSC}$^{1}$Department of Physics, University of California, Santa Cruz
 1156 High Street, Santa Cruz, CA 95064
}%
\affiliation{
\hypertarget{inst:SCIPP}$^{2}$Santa Cruz Institute for Particle Physics
1156 High Street, Santa Cruz, CA 95064
}
\affiliation{
    \hypertarget{inst:UCLA}$^{3}$Department of Physics \& Astronomy, University of California, Los Angeles, 430 Portola Plaza, Los Angeles, CA 90095, USA
}
\affiliation{
    \hypertarget{inst:OXF}$^{4}$University of Oxford, Wellington Square, Oxford OX1 2JD, UK
}
\affiliation{
\hypertarget{inst:USF}$^{5}$Department of Physics \& Astronomy, University of San Francisco, San Francisco, CA 94117, USA
}
\affiliation{
\hypertarget{inst:LBNL}$^{6}$Physics Division, Lawrence Berkeley National Laboratory, 1 Cyclotron Road, Berkeley, CA 94720, USA
}
\affiliation{
\hypertarget{inst:UCD}$^{7}$Department of Physics and Astronomy, University of California, Davis, 1 Shields Avenue, Davis, CA 95616, USA
}
\affiliation{
\hypertarget{inst:CASS}$^{8}$Centre for Astrophysics and Supercomputing, Swinburne University of Technology, PO Box 218, Hawthorn, VIC 3122, Australia
}
\affiliation{
\hypertarget{inst:ARC}$^{9}$The ARC Centre of Excellence for All Sky Astrophysics in 3 Dimensions (ASTRO 3D), Australia
}
\affiliation{
\hypertarget{inst:CFA}$^{10}$Center for Astrophysics, Harvard \& Smithsonian, Cambridge, MA 02138, USA
}
\affiliation{
\hypertarget{inst:NOIR}$^{11}$Gemini Observatory / NSF’s NOIRLab, Casilla 603, La Serena, Chile
}
\affiliation{
\hypertarget{inst:ESO}$^{12}$European Southern Observatory, Alonso de Cordova 3107, Casilla 19001, Vitacura, Santiago 19, Chile
}
\affiliation{
\hypertarget{inst:CAS}$^{13}$Chinese Academy of Sciences South America Center for Astronomy, National Astronomical Observatories, CAS, Beijing 100101, People's Republic of China
}
\affiliation{
\hypertarget{inst:IFA}$^{14}$Institute for Astronomy, University of Hawai'i, 2680 Woodlawn Drive, Honolulu, HI 96822-1897, USA
}
\affiliation{
\hypertarget{inst:SYD}$^{15}$Sydney Institute for Astronomy, School of Physics, A28, The University of Sydney, NSW 2006, Australia
}
\affiliation{
\hypertarget{inst:TUM}$^{16}$Technical University of Munich, TUM School of Natural Sciences, Physics Department, James-Franck-Straße 1, 85748 Garching, Germany
}
\affiliation{
\hypertarget{inst:MPA}$^{17}$Max-Planck-Institut für Astrophysik, Karl-Schwarzschild Straße 1, 85748 Garching, Germany
}
\affiliation{
\hypertarget{inst:NSW}$^{18}$University of New South Wales, Sydney, NSW 2052, Australia
}
\affiliation{
\hypertarget{inst:CMU}$^{19}$Department of Physics, McWilliams Center for Cosmology and Astrophysics, Carnegie Mellon University, 5000 Forbes Avenue, Pittsburgh, PA 15213, USA
}
\affiliation{
\hypertarget{inst:UChi}$^{20}$Department of Astronomy \& Astrophysics, University of Chicago, Chicago, IL 60637, USA
}
\affiliation{
\hypertarget{inst:KICP}$^{21}$Kavli Institute for Cosmological Physics, University of Chicago, Chicago, IL 60637, USA
}
\affiliation{
\hypertarget{inst:IUD}$^{22}$Center for Astronomy, Space Science and Astrophysics, Independent University, Bangladesh, Dhaka 1229, Bangladesh
}
\affiliation{
\hypertarget{inst:CO}$^{23}$The Observatories of the Carnegie Institution for Science, 813 Santa Barbara Street, Pasadena, CA 91101, USA
}

\date{\today}

\begin{abstract}
    We present a spectroscopic survey of field galaxies and lensed sources in the vicinity of the strong lensing galaxy cluster known as the Carousel lens at z=0.49. Using both Gemini/GMOS slitmask spectra and deep VLT/MUSE observations, we bring the total number of lensed sources up to 12, including 
    three which were not previously known from imaging observations but are apparent in the MUSE data as emission-line sources.  Of these sources, 10 have confident redshifts, and an additional 2 have tentative redshifts from likely Ly$\alpha$ emission (including seven new redshifts determined here adding to those presented previously in \cite{Sheu.Cikota.ea2024}).
    %measure 
    % These are 6-9,11-13
    %redshifts of 7 additional lensed sources,
    % These are 11,12,13
    %3 of which were not previously known or suspected from imaging observations.
    % JOD: The last w/o redshift is 10
    % JOD: My accounting is that LAEs 8 and 12 are confident, and 11 and 13 are tentative.
    % JOD: (Personally I am confident in 11 and 13, but their spectra are much less clear)
    %This brings the total number of lensed sources in the Carousel system to 13, of which 10 have confident redshifts, and an additional 2 have tentative redshifts from apparent Ly$\alpha$ emission.
    % JOD: Yes checked this
    The lensed sources span a redshift range from z=0.96 to 4.09 with most of them showing 3-5 images, including four sources displaying central or radial images.
    % JOD: Checked this as well, all good
    In total, we identify 42 images of these 12 sources.
    This lens system is remarkably symmetric and well-modeled by a simpler lens model than typical cluster lenses, and the large number of sources and their large range of redshifts make this cluster ideal for constraining cosmological parameters such as $w$ and $\Omega_m$ as well as the cluster density profile.
    Additionally, we present a catalog of 57 unlensed field galaxies with confident redshifts, of which 49 are associated with the cluster. We measure a cluster velocity dispersion of about 1100 km s$^{-1}$ from which we estimate a halo mass $M_{200c} \sim 1.2 \times 10^{15} M_\odot$.
\end{abstract}

\maketitle

\section{\label{sec:intro}Introduction}

Strong gravitational lenses with multiple source planes are potentially powerful probes of cosmology.  The ratio of lensing strength between each source plane constrains the dark energy equation of state and matter density independent of the Hubble constant \citep[e.g.][]{Gilmore.Natarajan2009,Collett2012}.  Strong lensing galaxy clusters are particularly powerful cosmic telescopes that can feature upward of tens of lensed sources, and cosmological constraints have been derived from a few multi-source plane lensing clusters \citep[e.g.][]{Jullo2010,Caminha2016,Caminha2022} and a group-scale system \citep{Bolamperti2024}.  In particular, \cite{Caminha2022} use five strong lensing clusters and show that combined with cosmic microwave background (CMB) data these improve constraints on $\Omega_M$ and $w$. Galaxy-scale double-source plane lenses can enable similar measurements \citep{Collett2014,Sahu2025,Bowden2025}, but require much larger samples to attain similar constraining power.  Cluster lenses have the advantage of a large number of sources, which give several cosmic baselines and break degeneracies in the mass model, but at the expense of more complicated mass models that e.g. must include cluster substructure.

We have identified an ideal, powerful, multi-source plane lensing cluster hosting more than 10 sources while having a very simple mass model \citep{Sheu.Cikota.ea2024}: the Carousel lens.  This lensing cluster was independently identified in Dark Energy Spectroscopic Instrument (DESI) Legacy Imaging \citep{Huang2021} and in the Dark Energy Survey \citep{O'Donnell2022}.  Hubble Space Telescope (HST) imaging (HST proposal \#16773; K. Glazebrook) revealed 10 probable lensed sources of which redshifts for five were determined from shallow MUSE data (Prog.~ID 0111.B-0400(H)) by \cite{Sheu.Cikota.ea2024}. \cite{Sheu.Cikota.ea2024} (Paper 0) also presented an initial lens modeling of this system, showing it to be well modeled by a simple, two-component mass model.  A companion paper, Paper II \citep{PaperII}, presents an updated lens modeling and cosmological constraints using four source planes.  %Even considering systematic uncertainties in the lens model and this limited number of sources, the Carousel lens gives strong constraints on cosmology.

In this paper, we present the results of new spectroscopic follow-up of this cluster, including deeper VLT MUSE data and Gemini GMOS targeted spectroscopy.  These data allow us to identify new lensed sources, bringing the total to 12, and to determine the redshifts for additional sources, giving firm redshifts for 9 and probable redshifts for 2 more.\footnote{One of the lensed sources identified in Paper 0 occurs in only a single image, and is therefore not a true strongly-lensed source. For this reason, the numbering of sources extends up to 13.}  We also determine the redshifts for 57 unlensed galaxies in the cluster field, of which 49 are cluster members.

The rest of the paper is structured as follows. In Section 2, we discuss the spectroscopic data and catalogs used.  Section 3.1 presents analysis results for the lensed sources, while 3.2 presents redshift determinations for unlensed field and cluster galaxies.  The cluster member catalog is used to model the cluster velocity dispersion in 3.3.

\section{\label{sec:data}Data}

\begin{figure}[!h]
    \centering
    \includegraphics[width=\linewidth]{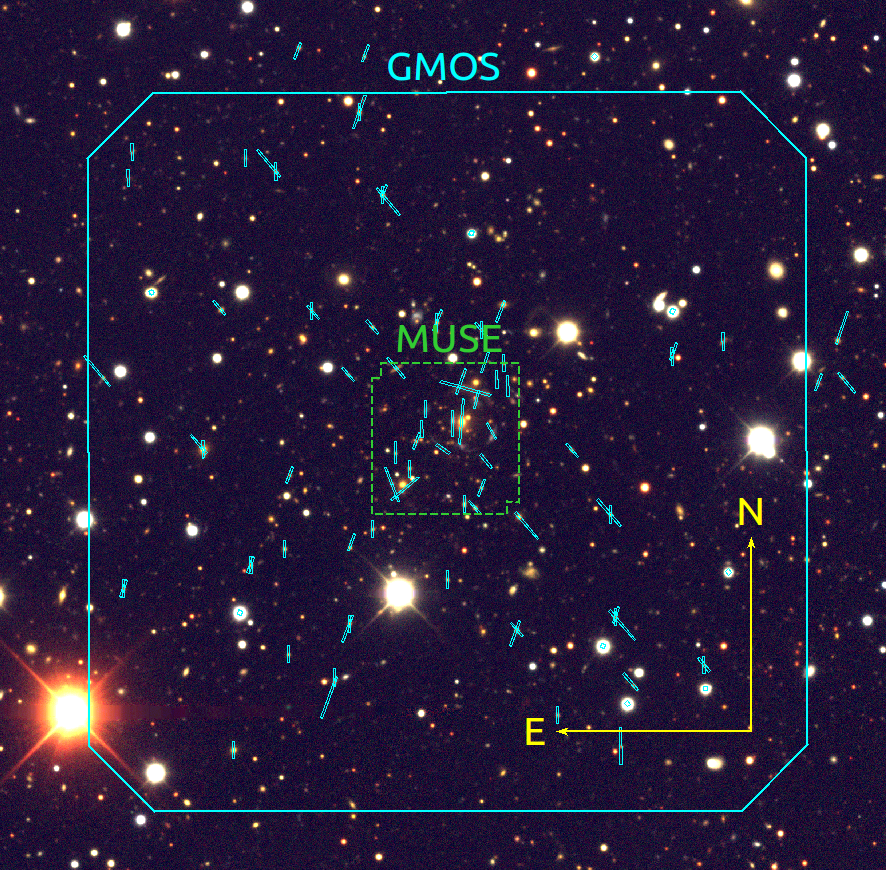}
    \caption{\textbf{Spectral FOV Comparison:} A comparison of the fields-of-view of GMOS slitmasks and MUSE observations used in this work. The slits of all four GMOS slitmasks are shown in cyan. The broader cyan region shows the size of the GMOS slitmask for scale, not the actual orientation used in our four masks. The compass arrows in the bottom-right are each 1.5' long.}
    \label{fig:spectra-fov}
\end{figure}

We summarize spectroscopic observations used in this work below. A comparison of the FOV of both VLT/MUSE and Gemini/GMOS observations can be found in Figure~\ref{fig:spectra-fov}. This work additionally makes use of HST WFC3 imaging in the F200LP and F140W bands (Prog. ID. 16773, PI Glazebook), and Dark Energy Camera Legacy Survey (DECaLS) Data Release 10 object catalogs \citep{LegacySurvey}.

\subsection{\label{sec:data_muse}VLT/MUSE Spectroscopy}

The Carousel lens system was observed with deep VLT/MUSE \citep{Bacon.Accardo.ea2010} observations in December 2024 and January 2025 (Prog. ID 0114.A-2018(A), PI Barone). These observations were attained in MUSE Wide-Field Mode (WFM) with adaptive optics (AO), with two overlapping pointings offset by approximately 8'' and covering a combined area of approximately 1.28 square arcmin, covering a wavelength range of approximately 470-935nm with a spectral resolution ranging between $\sim1600-3500$. Both pointings received eight exposures of 620s each, for a total exposure of 2.76 hours. The combined region covered by these observations is shown in Fig.~\ref{fig:spectra-fov}. These observations were reduced and coadded into a single data cube using the MUSE pipeline version 2.10.10 \cite{muse-pipeline}. Additional post processing improved sky subtraction residuals with \texttt{ZAP} \citep{MUSEZAP}, and corrected astrometry to align with two stars in the GAIA DR3 catalog \citep{GAIA-DR3} in the South-East of the field.
% These stars are 2885999499272085376 and 2885999494977103872 but that's a mouthful
The final data cube used in this work has a PSF of approximately 0.7'' FWHM.
 % JOD: I just took one of the GAIA stars and fit a 2d Gaussian to it. That seemed fine and gave sigma = 0.3", FWHM=0.7".

%\begin{figure}[!h]
%    \centering
    %\includegraphics[width=0.9\linewidth]{figs/HST_MUSE_FOV.png}
%    \includegraphics[width=0.9\linewidth]{figs/MUSE_FOV.png}
%    \caption{\textbf{MUSE data:} The reduced MUSE data described in Section~\ref{sec:data_muse}. The full data cube has been binned into three (red, green, blue) channels, representing ($>0.8\mu$m, 0.6-0.8$\mu$m, $< 0.6\mu$m) respectively, and displayed with a log-stretch.}
%    \label{fig:pretty_image}
%\end{figure}

\subsection{\label{sec:data_gmos}Gemini/GMOS Spectroscopy}

The Carousel system was the target of Gemini/GMOS observations with several goals: (i) identifying the redshifts of strongly lensed sources, (ii) measuring redshifts of likely cluster member galaxies, and (iii) measuring stellar velocity dispersions of the cluster member galaxies. As part of Gemini program GS-2023B-Q-103, these data were taken in January 2024.

Gemini/GMOS observations consisted of four slit masks, one targeting the central galaxy and faintest arcs, and three targeting other lensed sources. Candidate member galaxies were selected from DES catalogs with a simple color-magnitude cut, and were included in all masks as secondary targets. All masks used the R400 grating, which provided an ideal balance of wavelength coverage and spectral resolution for both identifying redshifts and measuring stellar kinematics. These observations span a wavelength range of approximately 550-1050nm, at a spectral resolution of roughly R $\sim 2000$. A summary of these observations, including exposure times, the number of candidate cluster members observed, and a list of lensed sources targeted by each mask, is given in Table~\ref{tab:gemini-gmos}. Note that some candidate member galaxies were observed on multiple masks.

% JOD: 
% BCG mask: Central wavelngth 720 +- 10nm, 9 exposures of 20 min, 3h total
% All other masks: Central wavelength 820 +- 10nm, 3 exposures 20min, ,1h total
% All: 4x1 binning (spatial x spectral)
% Filter? BCG: GG455_G0329, other: OG515_G0330

Gemini/GMOS data were reduced using \PypeIt\, \citep{PypeIt_JOSS,PypeIt_Zenodo},
a Python package for processing slit spectra from a wide array of spectrographs. As part of this work, \PypeIt\,was extended to correctly handle Gemini/GMOS data with tilted slits, functionality which is now available since \PypeIt\, version 1.18. These data were reduced using a development version of \PypeIt\, 1.17.4.

\section{\label{sec:results}Results}

\begin{table*}[]
    \renewcommand{\arraystretch}{1.2}
    \centering
    \begin{tabular}{c|c|c|c|c|c}
     \multirow{2}*{Mask} & Grating & Total & $\lambda_{\text{cen}}$  & Field & Lensed \\
     & and Filter & Exposure & (nm) & Galaxies & Sources \\
     \hline
    \texttt{bcg-01} & R400, GG455 & 10800s & 710-730 & 19* & 3a,8b,9a \\
    \texttt{arcs-11} & R400, OG515 & 3600s & 780-800 & 19 & 3a-c,4c,5b,7c \\
    \texttt{arcs-12} & R400, OG515 & 3600s & 810-830 & 16 & 4a,7a,9c,10\\
    \texttt{arcs-13} & R400, OG515 & 3600s & 790-810 & 14 & 1a,4b,8a \\
    \end{tabular}
    \caption{\textbf{GMOS observations:} Summary of Gemini/GMOS observations. *Excluding BCG}
    \label{tab:gemini-gmos}
\end{table*}

\begin{figure*}[ht!]
    \centering
    \includegraphics[width=1.0\linewidth]{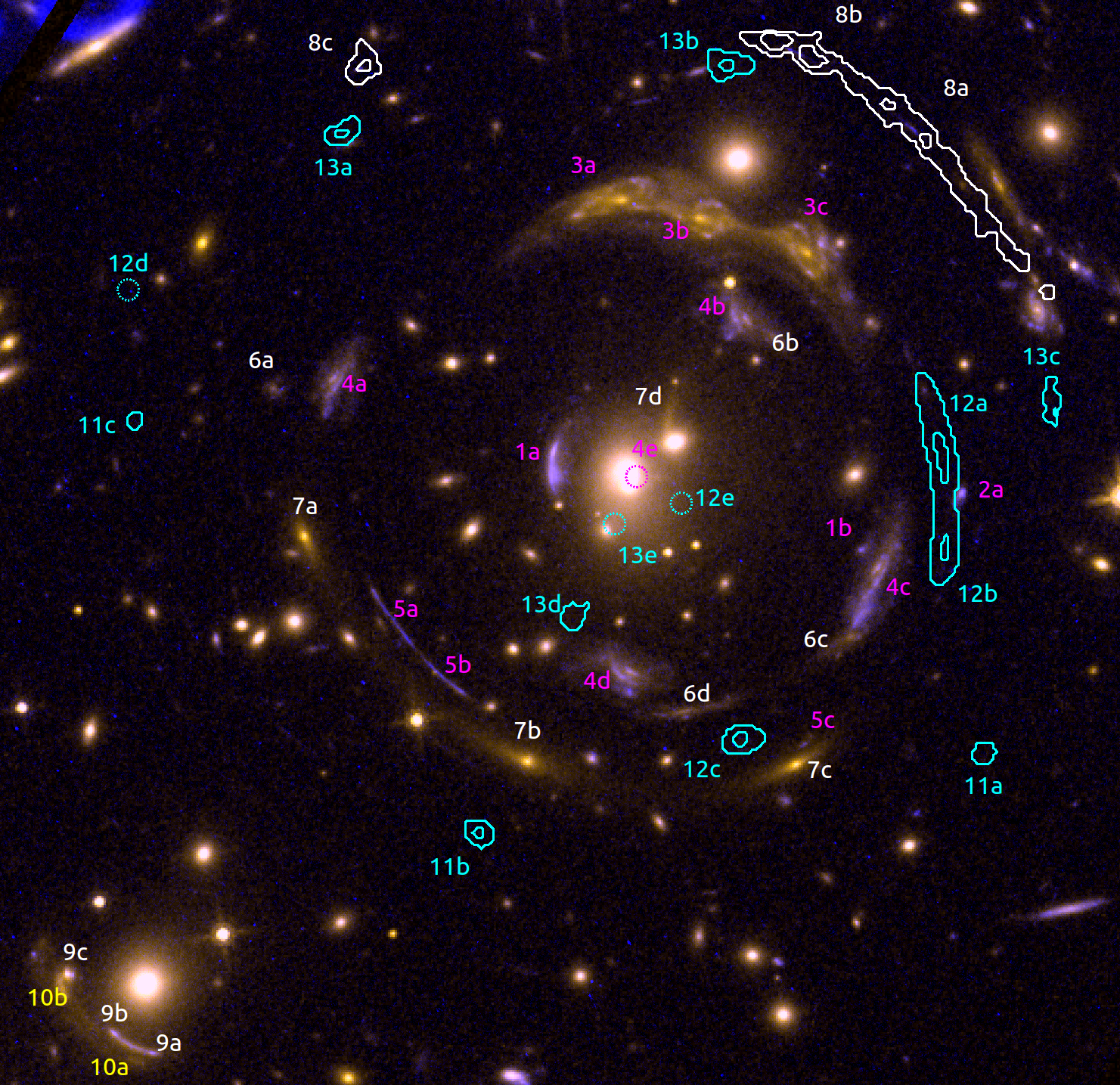}
    \caption{\textbf{All Sources}: All known lensed sources in the Carousel system, overlaid on HST imaging with color coding as follows. Magenta: Known sources with redshifts presented in Paper 0 \citep{Sheu.Cikota.ea2024}; White: Previously known sources with redshifts presented in this work; Cyan: Newly discovered sources, with extended emission observed in MUSE data; Yellow: Sources which do not yet have confident redshifts. Sources 8, 11, 12, and 13 display contours of Ly$\alpha$ emission.}
    \label{fig:all-sources}
\end{figure*}

Here we present the results of the spectroscopic modeling of both lensed and unlensed objects in the Carousel field. Strongly lensed sources are described in Section~\ref{sec:results_sources}. Spectra and redshifts of field galaxies, including cluster members, are presented in Section~\ref{sec:results_galaxies}. This catalog is then used to study the dynamical state of the Carousel cluster in Section~\ref{sec:results_vdisp}.
All redshifts were obtained using \MARZ\, \citep{MARZ} unless otherwise noted.

\subsection{\label{sec:results_sources}Strongly Lensed Sources}

% JOD: The Carousel got ~47:00 observations on two separate nights in September 2023, under two separate program IDs, 0109.B-0871(D) and 0111.B-0400(H).
% JOD: However, it appears the former had especially bad weather, so William's paper only used the latter.
Previously, \cite{Sheu.Cikota.ea2024} identified sources 1 through 7 behind the Carousel cluster, and reported confident redshifts for all except Sources 6 and 7 using MUSE observations with a total integration time of $\approx47$ minutes (Prog. ID 0111.B-0400(H), PI Bian). Combining HST imaging and newer spectroscopic follow-up, including deeper MUSE observations, we report 6 further objects strongly lensed by the Carousel, bringing the sample of known multiply imaged sources up to 13\footnote{Source 2 is only singly imaged, and we therefore do not consider it to be a `true' strongly lensed source. It is included here for consistency with Paper 0.}. We continue the labeling scheme introduced in \cite{Sheu.Cikota.ea2024}, identifying these new sources as 8 through 13. Figure~\ref{fig:all-sources} displays all lensed sources known to date, color coded to denote previously known or newly discovered sources, with and without confident redshifts.

In addition to extracting spectra of known objects, we perform a search for other emission line sources in the MUSE data described in Section~\ref{sec:data_muse}. This included stepping through each wavelength slice of the data cube using a number of settings, looking at both individual slices and binning by 10 spectral pixels, as well as continuum subtracting from nearby wavelength ranges to explicitly look for emission. As a result of these searches, we identify clear emission associated with newly identified strongly lensed sources 11, 12, and 13, all of which exhibit Lyman $\alpha$ emission at redshift $z>3$.

In this section we describe the search for redshifts of strongly lensed sources. The MUSE and GMOS spectra of sources previously identified in imaging data are described in Section~\ref{sec:results_known_sources}. Four Lyman $\alpha$ emitting sources are described in Section~\ref{sec:results_LAEs}. %A full list of all sources, including those previously presented in Paper 0, can be found in the appendix, along with spectra of all sources with newly reported redshifts.
We present a full catalog of lensed sources in the field, including those previously reported in Paper 0 \citep{Sheu.Cikota.ea2024}, in Table~\ref{tab:full-source-list}. Table~\ref{tab:full-source-list} gives the name, position, and redshift of each strongly-lensed source, as well as spectral features from which the redshifts were derived. Note that coordinates of each source are provided for clarity, and are not intended as precision measurements.

%We additionally present the MUSE and GMOS spectra of lensed sources with newly published redshifts. Sources previously identified from imaging data are shown in Figure~\ref{fig:source-spectra}, and the four LAEs are presented in Figure~\ref{fig:LAE-spectra}.

\begin{table*}[h!]
    \renewcommand{\arraystretch}{1.2}
    \centering
    \begin{tabular}{ccccc}
        \toprule Source Label & R.A. & Dec. & Redshift & Spectral Features \\
        \midrule
\textit{1a} & 06:03:56.86 & -35:58:05.05 & \textit{0.962} & See Paper 0 \\
\textit{1b} & 06:03:55.42 & -35:58:09.87 & --- & --- \\[.3em]

\textit{2a}$^\dagger$ & 06:03:54.95 & -35:58:06.65 & \textit{0.962} & See Paper 0 \\[.3em]

\textit{3a} & 06:03:56.53 & -35:57:49.95 & \textit{1.166} & See Paper 0\\
\textit{3b} & 06:03:56.17 & -35:57:51.00 & --- & --- \\
\textit{3c} & 06:03:55.66 & -35:57:53.13 & --- & --- \\[.3em]

\textit{4a} & 06:03:57.88 & -35:58:00.16 & \textit{1.432} & See Paper 0\\
\textit{4b} & 06:03:55.96 & -35:57:56.76 & --- & --- \\
\textit{4c} & 06:03:55.34 & -35:58:10.92 & --- & --- \\
\textit{4d} & 06:03:56.50 & -35:58:16.77 & --- & --- \\
\textbf{4e} & 06:03:56.47 & -35:58:05.72 & \textbf{1.432} & [O II] \\[.3em]

\textit{5c} & 06:03:55.68 & -35:58:20.92 & \textit{1.432} & See Paper 0\\
\textit{5b} & 06:03:57.29 & -35:58:17.88 & --- & --- \\
\textit{5a} & 06:03:57.69 & -35:58:12.40 & --- & --- \\[.3em]

\textit{6d}* & 06:03:56.24 & -35:58:19.04 & \textbf{1.656} & Fe II, Mg II \\
\textit{6c}* & 06:03:55.46 & -35:58:14.74 & --- & --- \\
\textbf{6b} & 06:03:55.82 & -35:57:57.76 & --- & --- \\
\textbf{6a} & 06:03:58.17 & -35:58:00.65 & --- & --- \\[.3em]

\textit{7a} & 06:03:58.02 & -35:58:09.07 & \textbf{1.627} & C III], C II], \MARZ\,template fit \\
\textit{7b} & 06:03:56.98 & -35:58:21.89 & --- & --- \\
\textit{7c} & 06:03:55.72 & -35:58:22.05 & --- & --- \\
\textit{7d} & 06:03:56.32 & -35:58:02.53 & n.a. & Not firmly detected \\[.3em]

\textbf{8a} & 06:03:55.17 & -35:57:46.09 & \textbf{3.549} & Ly$\alpha$, O III], C III] \\
\textbf{8b} & 06:03:55.75 & -35:57:41.46 & --- & --- \\
\textbf{8c} & 06:03:57.76 & -35:57:42.51 & --- & --- \\[.3em]

\textbf{9a} & 06:03:58.73 & -35:58:38.46 & \textbf{1.507} & Fe II, Mg II, [O II] \\
\textbf{9b} & 06:03:58.91 & -35:58:37.33 & --- & --- \\
\textbf{9c} & 06:03:59.11 & -35:58:34.02 & --- & --- \\[.3em]

\textbf{10a} & 06:03:58.83 & -35:58:38.87 & n.a. & n.a. \\
\textbf{10b} & 06:03:59.15 & -35:58:34.87 & n.a. & n.a. \\[.3em]

% JOD: removed 'tentative' designation
%\textbf{11a} & 06:03:54.84 & -35:58:21.43 & \textbf{4.090$^\dagger$} & Ly$\alpha$, He II \\
\textbf{11a} & 06:03:54.84 & -35:58:21.43 & \textbf{4.090} & Ly$\alpha$, He II \\
\textbf{11b} & 06:03:57.20 & -35:58:26.02 & --- & --- \\
\textbf{11c} & 06:03:58.81 & -35:58:02.53 & --- & --- \\[.3em]

\textbf{12a} & 06:03:55.06 & -35:58:04.28 & \textbf{3.086} & Ly$\alpha$, Si II, C IV, O III], C III] \\
\textbf{12b} & 06:03:55.04 & -35:58:09.46 & --- & --- \\
\textbf{12c} & 06:03:55.98 & -35:58:20.76 & --- & --- \\
\textbf{12d} & 06:03:58.84 & -35:57:55.08 & --- & --- \\
\textbf{12e} & 06:03:56.26 & -35:58:07.23 & --- & --- \\[.3em]

% JOD: removed 'tentative' designation
%\textbf{13b} & 06:03:56.08 & -35:57:42.16 & \textbf{3.086$^\dagger$} & Ly$\alpha$ \\
\textbf{13b} & 06:03:56.08 & -35:57:42.16 & \textbf{3.086} & Ly$\alpha$ \\
\textbf{13a} & 06:03:57.83 & -35:57:45.90 & --- & --- \\
\textbf{13c} & 06:03:54.52 & -35:58:01.43 & --- & --- \\
\textbf{13d} & 06:03:56.76 & -35:58:13.56 & --- & --- \\
\textbf{13e} & 06:03:56.57 & -35:58:08.38 & --- & --- \\[.3em]
        
        \bottomrule
    \end{tabular}
    \caption{\textbf{Full Source List:} Full list of strongly-lensed sources. Source positions and redshifts previously reported in Paper 0 \citep{Sheu.Cikota.ea2024} are shown in italics, and those newly reported in this work are shown in bold. A dash (---) indicates the same value as the previous row. Sources or images without confirmed redshifts are listed as ``n.a.''. *Sources 6d and 6c were reported in Paper 0 with different labels. We have renamed them for consistency with source group 4. $\dagger$Source 2 occurs in only one image, and therefore is not a true strongly-lensed source. It is included here for consistency with Paper 0 \cite{Sheu.Cikota.ea2024}. %$\dagger$Redshifts of sources 11 and 13 are considered tentative, as they are based primarily on a single spectral feature.
    }
    \label{tab:full-source-list}
\end{table*}

\subsubsection{\label{sec:results_known_sources}Known Sources}

\newlength{\specboxsize}
\setlength{\specboxsize}{4.5cm}
\begin{figure*}[!h]
    \centering
    \begin{tabular}{ll}
    \centering
        \includegraphics[width=\specboxsize]{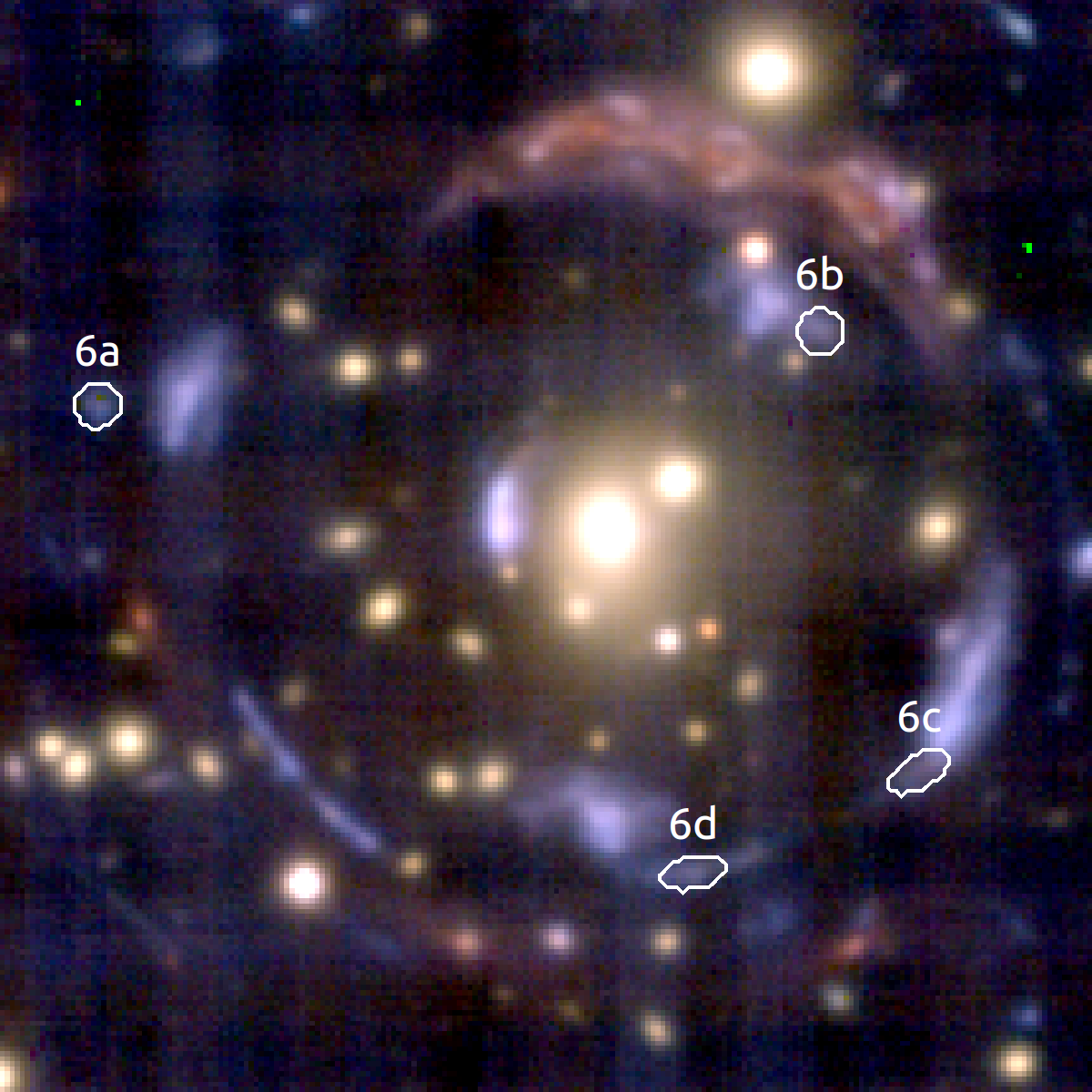} & \includegraphics[height=\specboxsize]{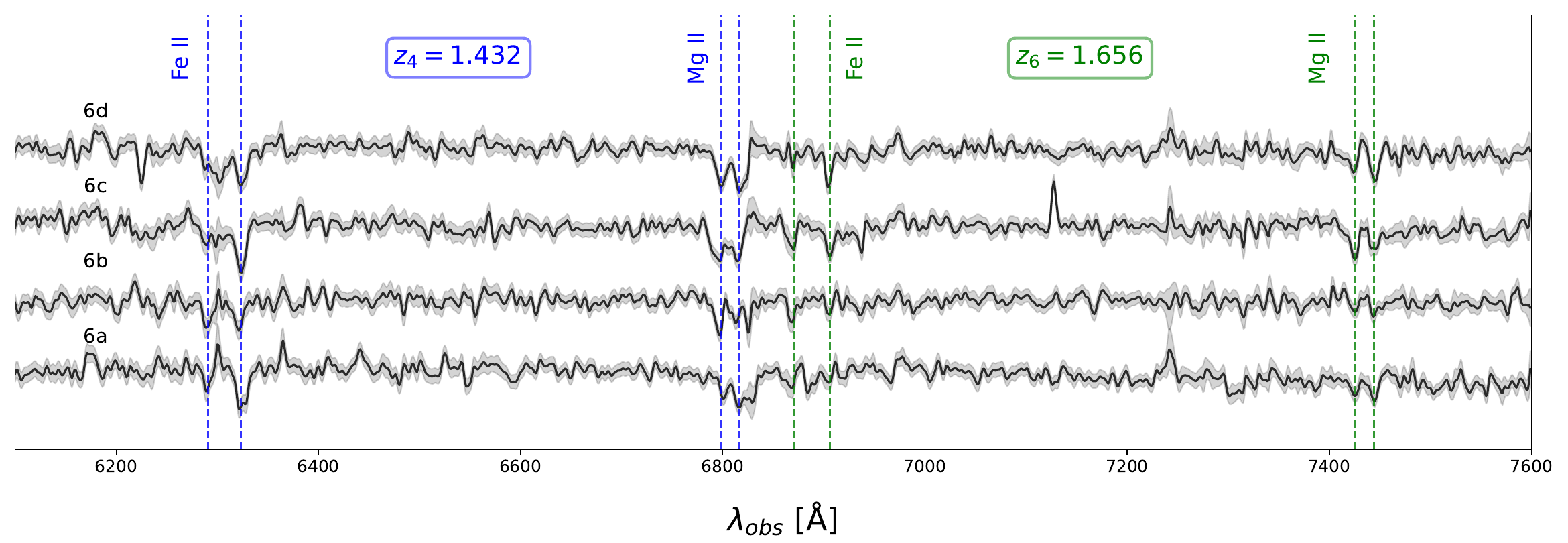} \\
        \includegraphics[width=\specboxsize]{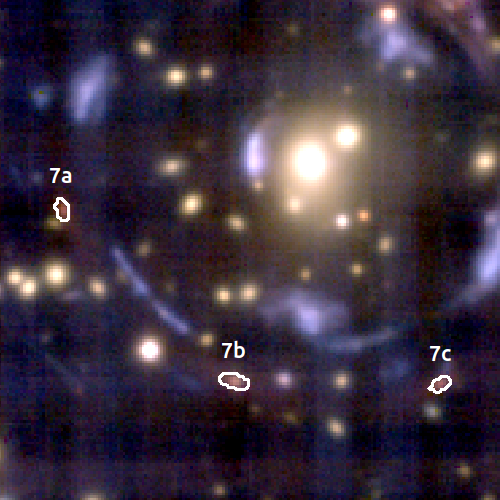} & %\includegraphics[height=\specboxsize]{figs/source_spectra/source_7_spec_2025-12-18.pdf}\\
        \includegraphics[height=\specboxsize]{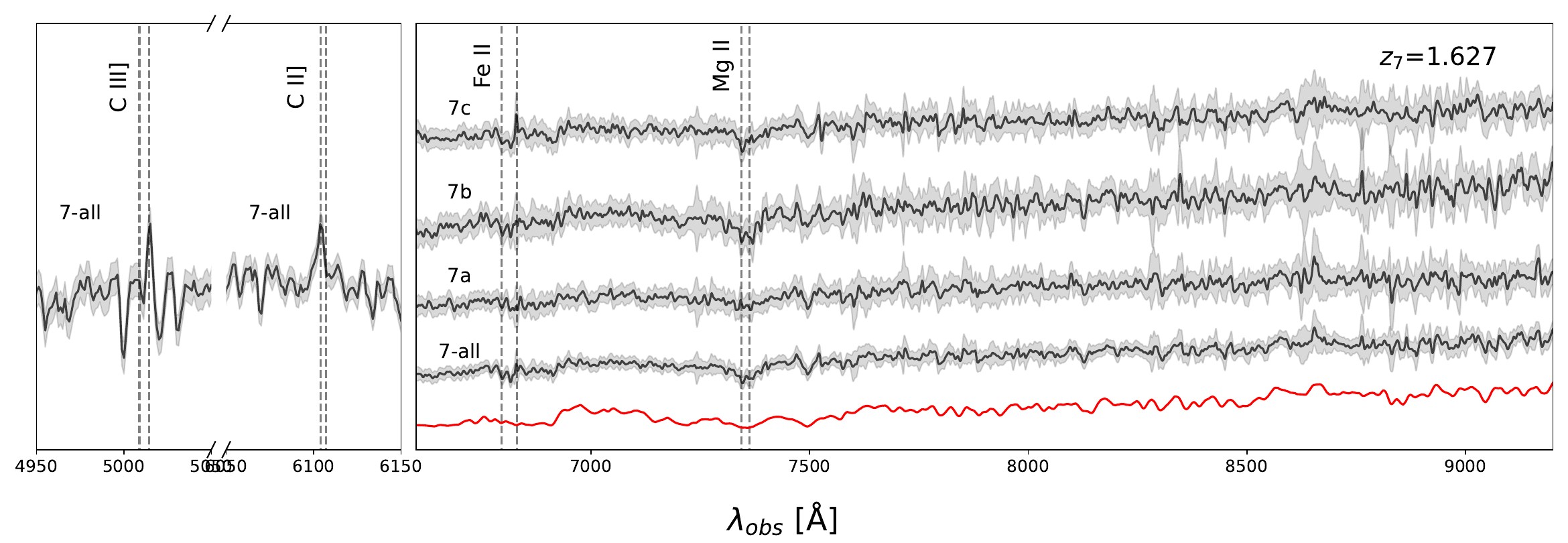} \\
        \includegraphics[width=\specboxsize]{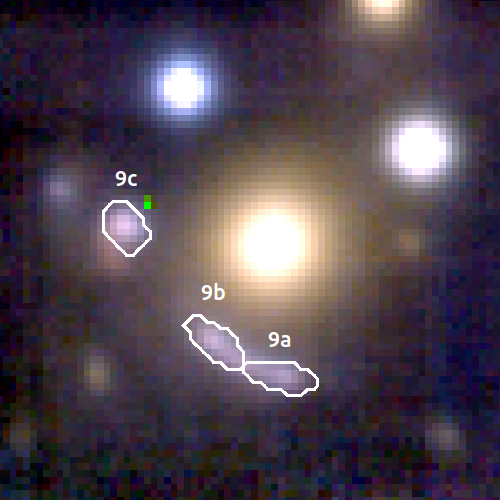}& 
        \includegraphics[height=\specboxsize]{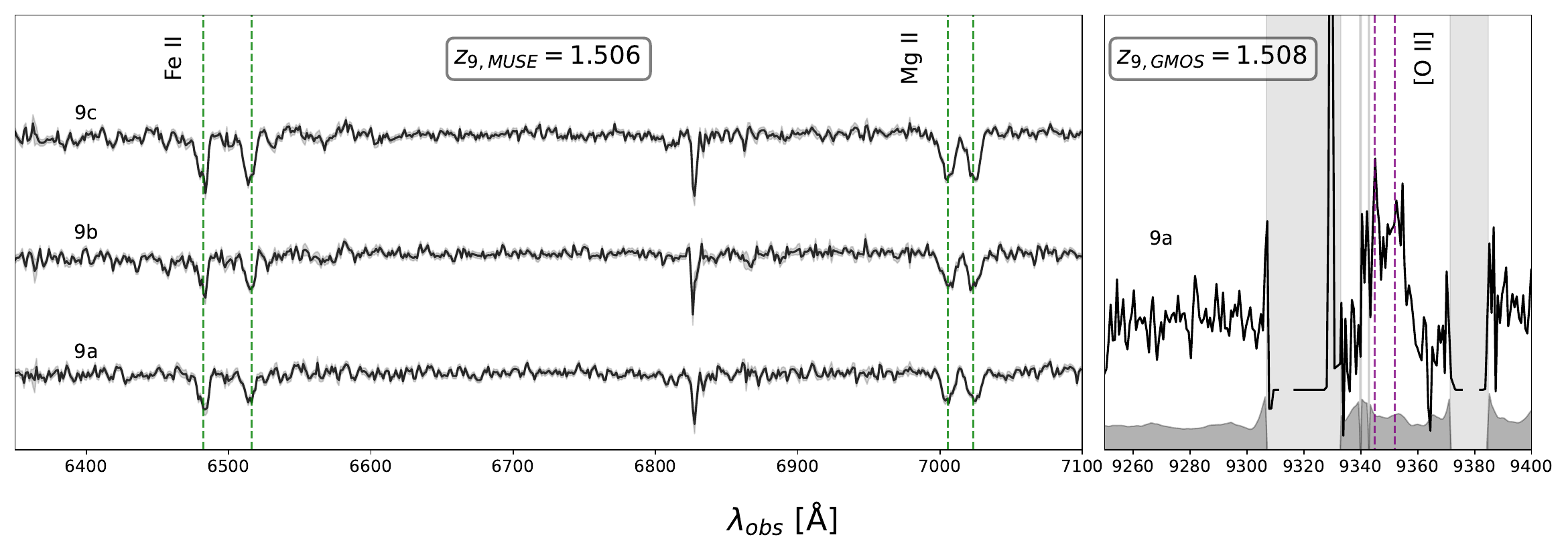} \\
    \end{tabular}

    \caption{\textbf{Known sources:} Spectra of sources previously identified in imaging data. \textbf{Left:} MUSE data, with regions of extracted spectra overlaid. \textbf{Right:} Extracted spectra of each lensed source, identifying spectral features and redshift. For clarity, the spectrum of each lensed image is offset vertically. For Source 7, we include the Early Type Galaxy \MARZ\, template match in red. For source 9, we additionally show a GMOS spectrum of image 9a exhibiting a clear [O II] doublet, with uncertainty bands shown at the bottom of the plot. Wavelengths contaminated by poor sky subtraction are indicated in vertical gray bands.}
    \label{fig:source-spectra}
\end{figure*}

Four images of Source 6 are apparent, in a configuration close to Source 4. Paper 0 reported two images of Source 6 (6c and 6d), and we confirm two additional images (6a and 6b). While Source 6 exhibits no emission lines, all four images have clear absorption features apparent in MUSE data. These absorption features match two distinct redshifts: Fe II 2587,2600\AA\, and Mg II 2796,2803\AA\, at both z=1.656 and z=1.432. We conclude that the true redshift is $z_6=1.656$, and each image of Source 6 additionally exhibits absorption from the CGM of Source 4, at $z_4=1.432$. This interpretation is consistent with a precise strong lens model (Paper II, \cite{PaperII}). Features for both redshifts are labeled for all four images of Source 6 in Figure~\ref{fig:source-spectra}.

While Paper 0 reported four images of Source 4, the newly available MUSE data demonstrates an additional image, a faint central image which we label 4e. Image 4e exhibits the same [O II] emission doublet as 4a-d. This [O II] emission doublet is also seen in the GMOS slit spectra of the central cluster galaxy, further confirming image 4e.

Source 7 similarly exhibits no emission features, but clear continuum in the near infrared. Template matching using \MARZ\,provided an excellent match to both the continuum shape and faint absorption features in 7a-c, from which two experts (TEJ and KG) independently inferred a redshift of z=1.627 with an `Early Type Absorption Galaxy' template.  The candidate radial image 7d, however, is both demagnified and blended with the light of a cluster galaxy, and therefore cannot be confirmed as the same object from spectroscopy alone. Ongoing modeling work suggests it is highly likely that 7d is in fact a counter image of Source 7.
% Originally we cited Paper II, but there is also ongoing work beyond Felipe's

Sources 9 and 10 were identified in HST imaging, and explicitly targeted with GMOS slit spectra. Both are lensed by a bright cluster member roughly 40'' from the Carousel BCG. Source 9 was assigned a tentative redshift from GMOS slit observations, which featured a candidate [O II] emission doublet. This redshift was confirmed by matching absorption features seen in MUSE, in all three images 9a,b,c. Source 10 is a faint red object appearing in several images just outside of Source 9; it could not be confidently distinguished from Source 9 in the MUSE cube. Source 10 is the only lensed source we identify without a redshift estimate.

\subsubsection{\label{sec:results_LAEs}Lyman $\alpha$ Emitters}

\begin{figure*}[!h]
    \centering
    \begin{tabular}{ll}
    \centering
        \includegraphics[width=\specboxsize]{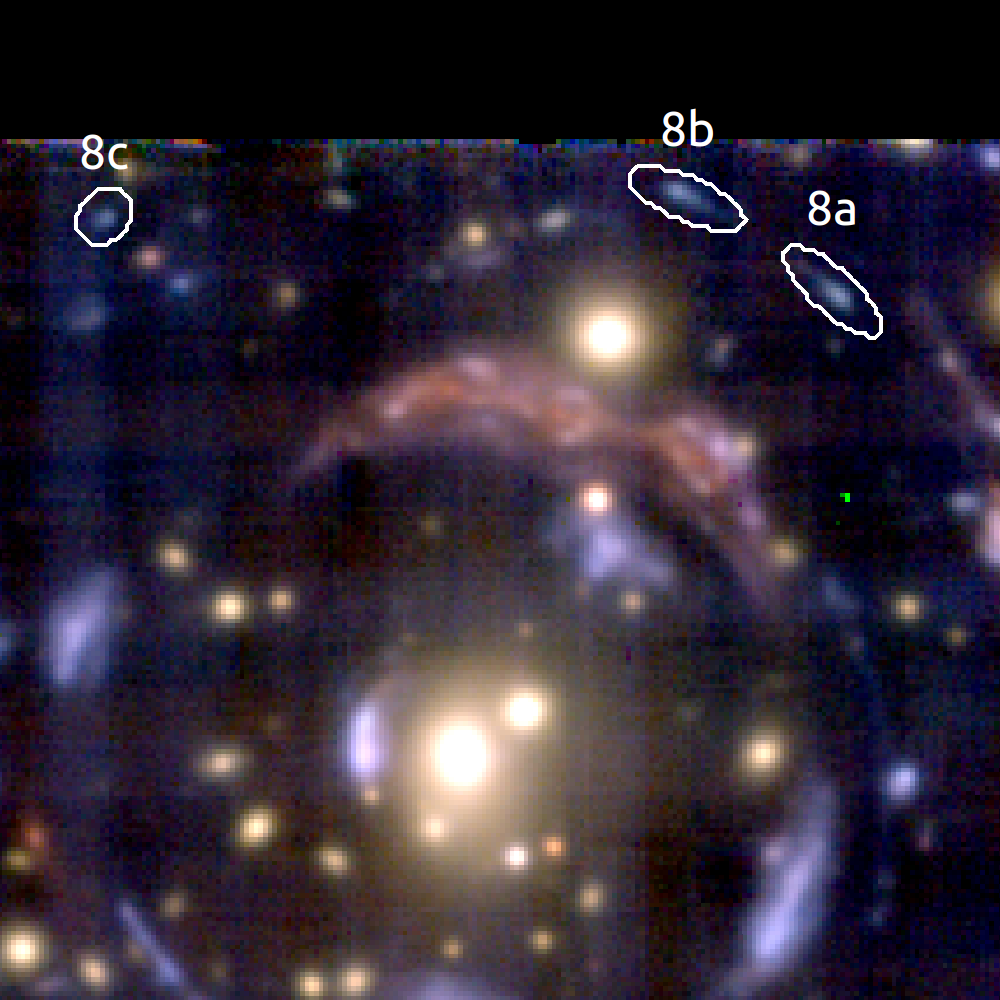}& \includegraphics[height=\specboxsize]{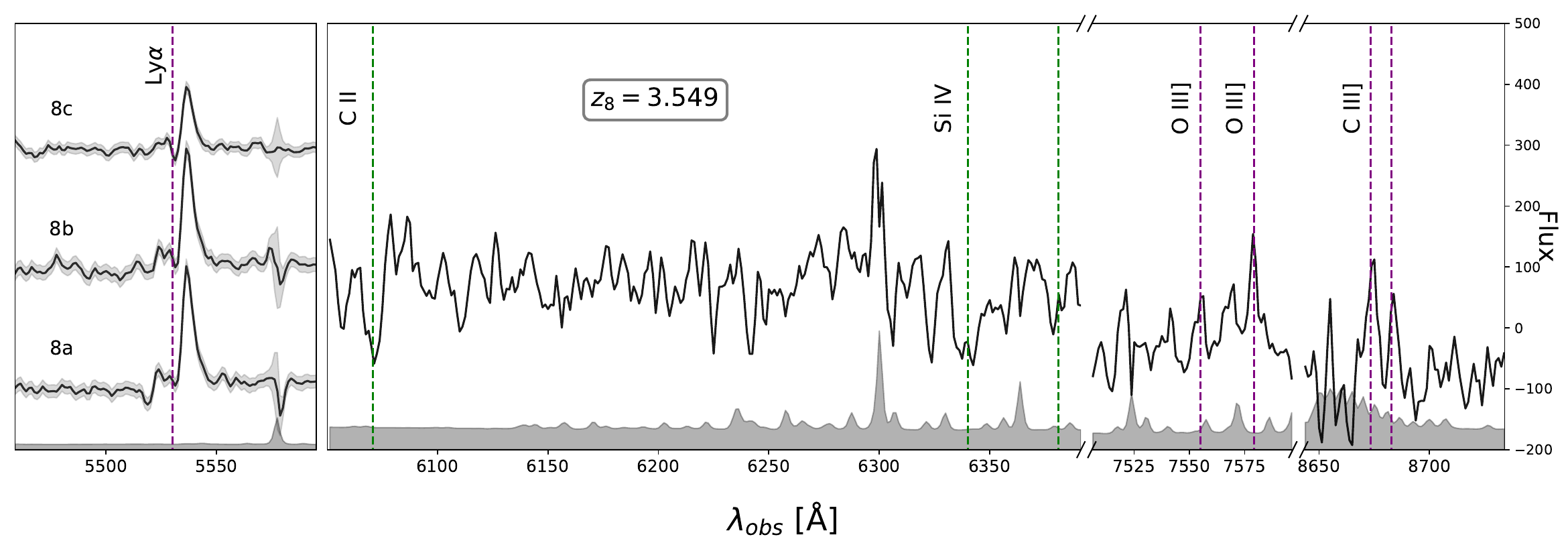} \\
        \includegraphics[width=\specboxsize]{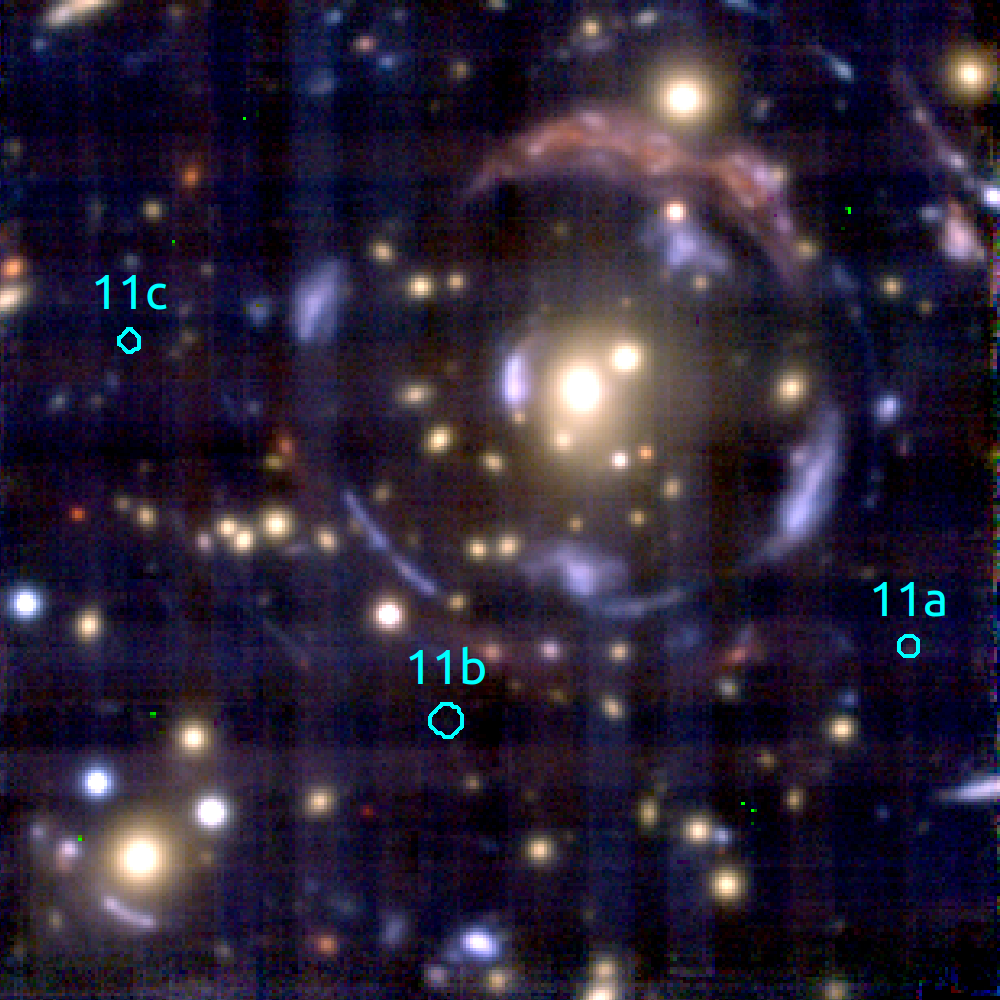}& \includegraphics[height=\specboxsize]{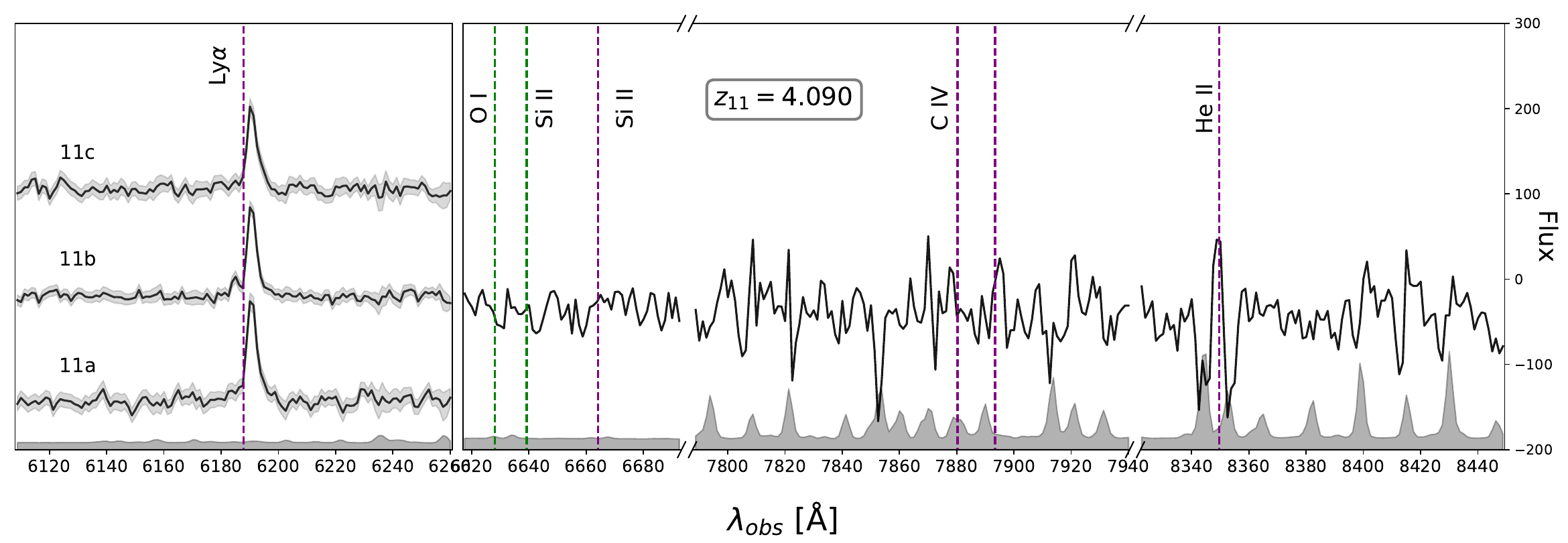} \\
        \includegraphics[width=\specboxsize]{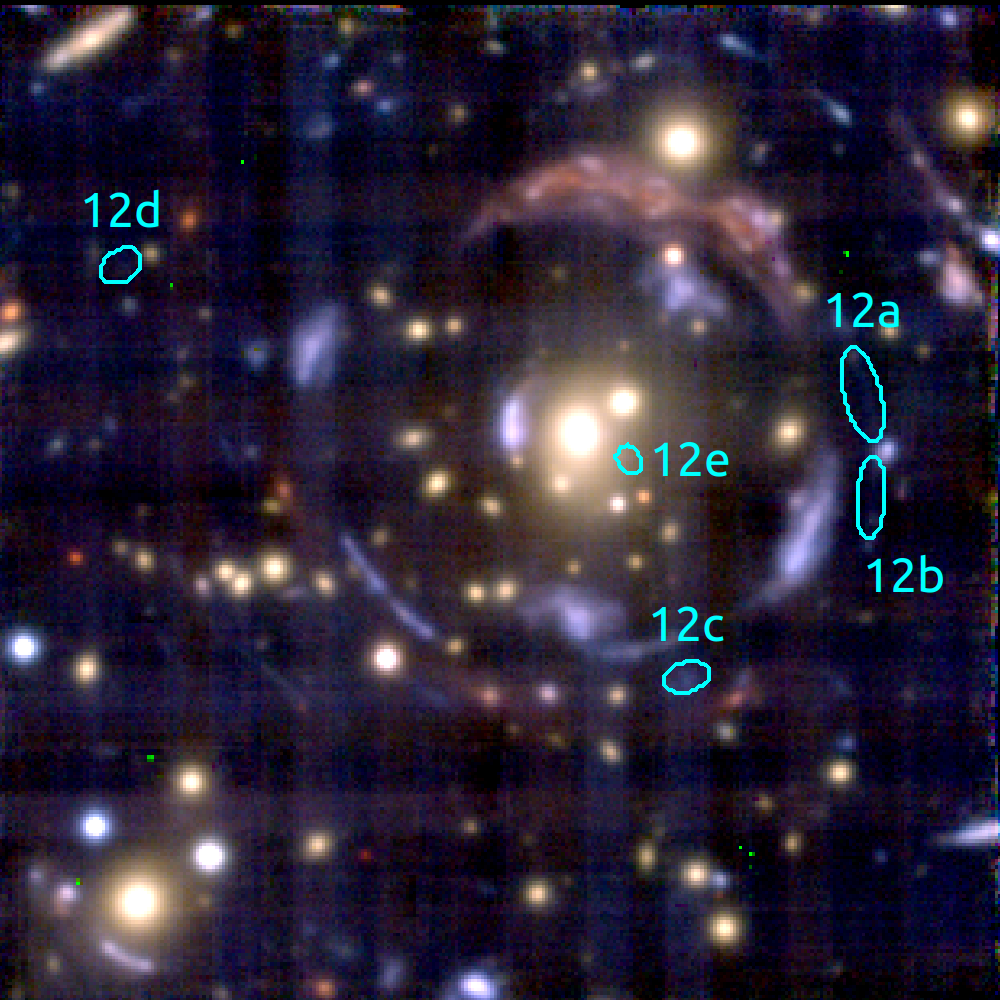}& \includegraphics[height=\specboxsize]{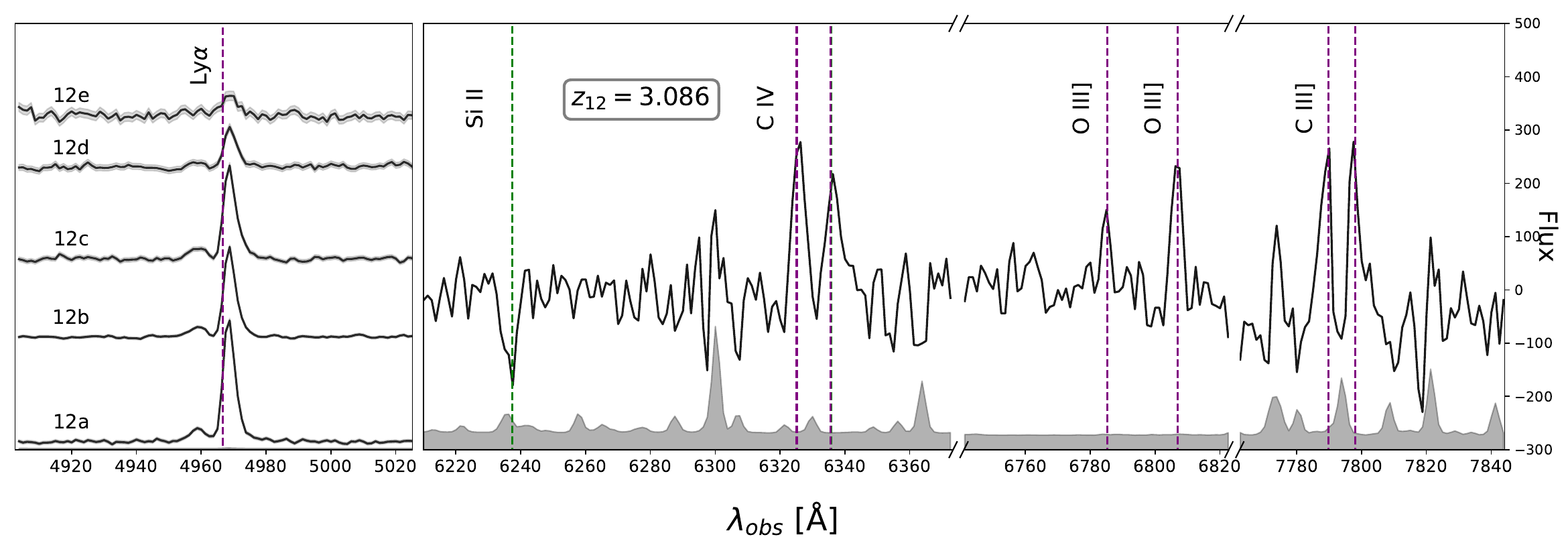} \\
        \includegraphics[width=\specboxsize]{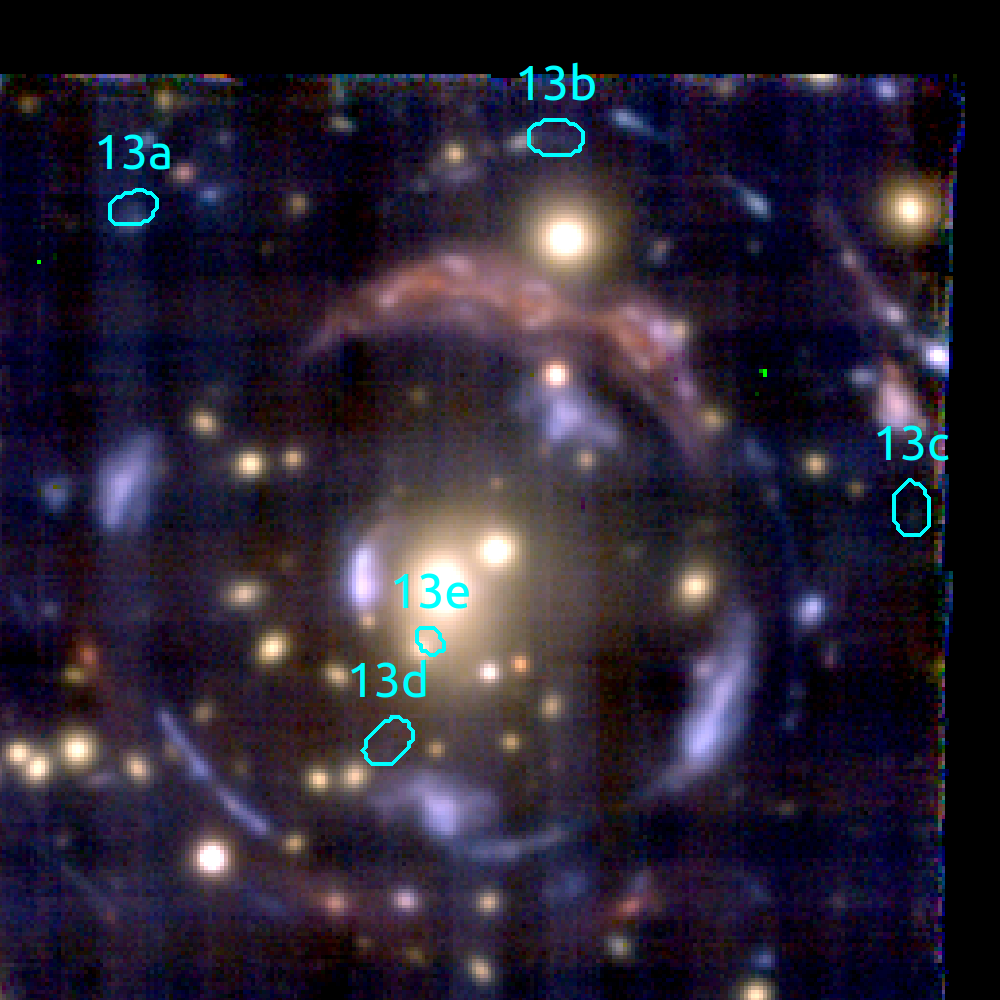}& \includegraphics[height=\specboxsize]{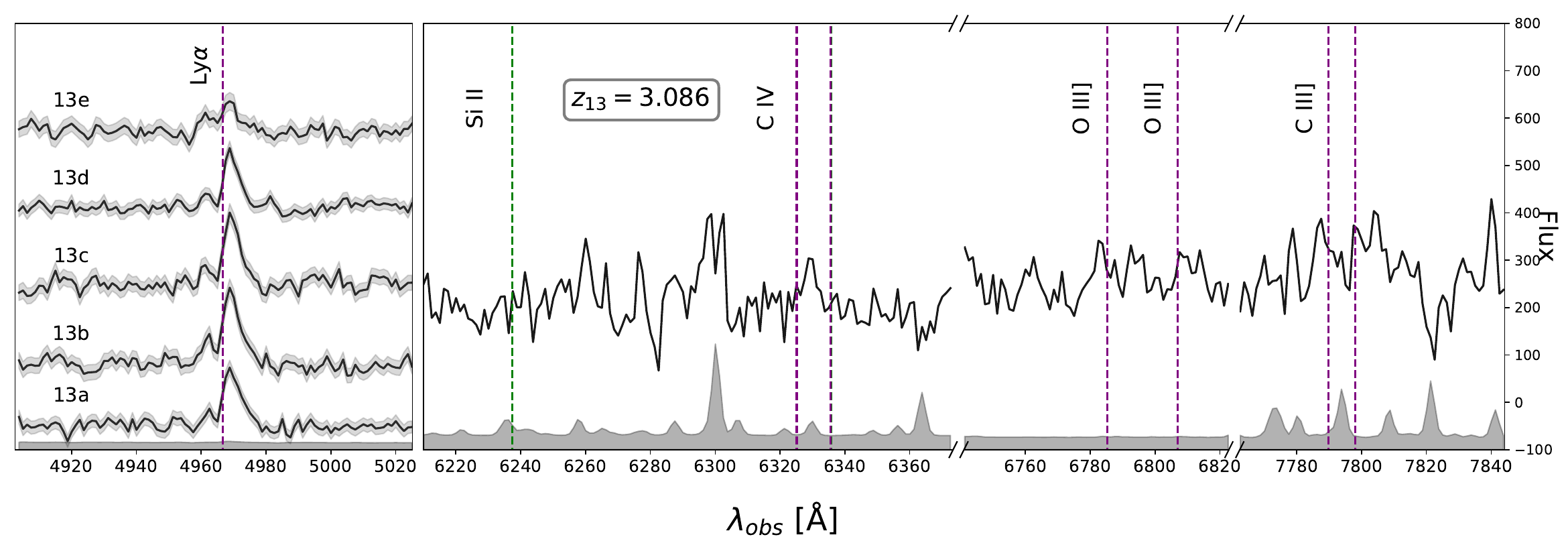} \\
    \end{tabular}

    \caption{\textbf{Lyman $\alpha$ Emitters:} Spectra and newly reported redshifts of strongly-lensed LAEs discovered in MUSE data. \textbf{Left Column:} RGB image of the MUSE data, with regions used for each object overlaid. \textbf{Right Column:} Extracted spectra of each object. In each plot, the left panel shows Ly$\alpha$ emission visible in each image, offset and scaled for clarity. The right panels show the stacked spectrum of all images, demonstrating faint features. Grey shaded regions at the bottom show the uncertainty. Features labeled in green (purple) are expected in absorption (emission).}
    \label{fig:LAE-spectra}
\end{figure*}

We identify four distinct strongly lensed Ly$\alpha$ emitters (LAEs) apparent in MUSE data, labeled sources 8 and 11-13. Only one of these, Source 8, was suspected from previous HST imaging data. All four LAEs exhibit the characteristic asymmetric double-peaked line profile of Lyman $\alpha$\footnote{See \cite{Hayes.Runnholm.ea2021} and \cite{Hayes.Runnholm.ea2023} for a detailed empirical overview of Ly$\alpha$ line profiles.}. We additionally search for faint rest-UV emission lines expected in weak Lyman break galaxies \citep{Shapley.Steidel.ea2003}. Contours of Ly$\alpha$ emission for all four LAEs are overlaid in Figure~\ref{fig:all-sources}, and MUSE spectra of all four LAEs can be found in the appendix.

Source 8 was initially identified as a pair of faint blue features in HST imaging, corresponding to images 8a and 8b. These images were targeted by GMOS observations, but could not be detected in GMOS data. In MUSE, these images were found to lie within an extended arc of emission, and a third image 8c was also identified. Source 8 additionally exhibits fainter rest-UV emission lines O III] 1661,1666\AA\, and C III] 1907,1909\AA\, in stacked spectra.
% Source 8 is faintly visible in HST, and also actually in the LS DR10 catalog
% 8a, 8b, 8c = (2185, 2288, 2659)

Source 11, in contrast to the other LAEs here, appears as a triplet of point sources rather than extended arcs. The only spectral feature observed in all three images of Source 11 is emission consistent with Ly$\alpha$ at $z_{11}=4.09$. While it is plausible that another emission double could be mistaken for Ly$\alpha$, the extreme asymmetry of flux between the red- and blue-shifted emission in Source 11 is unlikely to be explained by C IV 1548,1551 \AA, C III] 1907,1909\AA\, or [O II] 3726,3229\AA\, doublets which have similar spacing. The stacked spectrum of all three images further exhibits tentative He II 1640\AA\, emission consistent with $z_{11}=4.09$, and the three observed images of Source 11 are consistent with a strongly lensed source at this redshift given the lens model in Paper II \cite{PaperII}. We therefore consider it highly probable that Source 11 is a strongly lensed LAE at this redshift.

\begin{figure}[!h]
    \centering
    \includegraphics[trim={2.2cm 0cm 2.2cm 0cm},clip,width=1\linewidth]{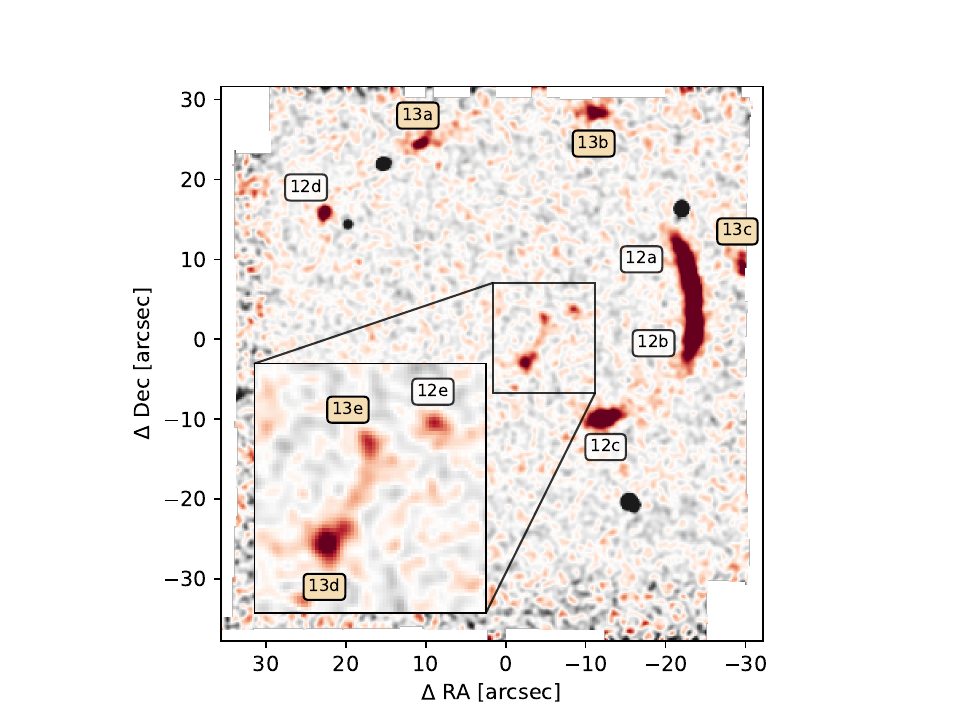}
    \caption{\textbf{Sources 12 and 13:} The spatial distribution of Ly$\alpha$ emission from sources 12 and 13, with each image labeled. Note that several ``hot pixels'' produce artifacts that appear as black dots. These artifacts do not contaminate any of the spectra extracted, and have no effect on the conclusions of this work.}
    \label{fig:sources-12-13}
\end{figure}

\begin{figure*}[!h]
    \centering
    \includegraphics[width=0.9\linewidth]{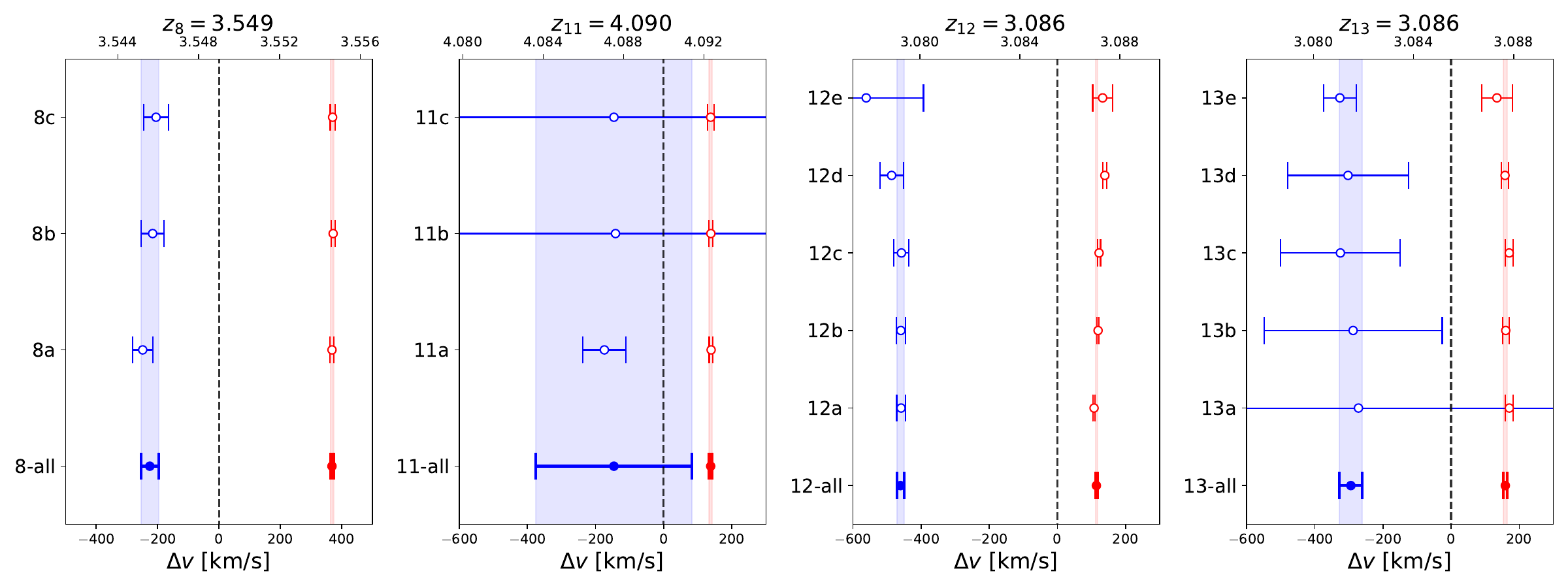}
    \caption{\textbf{LAE Kinematics:} Kinematics offsets of the red- and blue-shifted Ly$\alpha$ emission of all four Lyman $\alpha$ emitters, for each observed image. Kinematic offsets and uncertainties were computed with \texttt{ppxf} as described in Section~\ref{sec:results_LAEs}, using the same spatial regions of MUSE data as in Figure~\ref{fig:LAE-spectra}.}
    \label{fig:lya-kinematics}
\end{figure*}

Sources 12 and 13 appear at nearly identical redshift $z_{12} = z_{13} = 3.086$, forming 10 distinct images in total (see Figure~\ref{fig:all-sources}). The spatial distribution of Ly$\alpha$ flux at this redshift is shown in Figure~\ref{fig:sources-12-13}, with each observed position labeled. This image was produced by summing the continuum-subtracted flux between 4962 and 4972 \AA, and spatially convolving this flux with a Gaussian of $\sigma=1.5$px (approximately 0.3''). Initially, it was suspected that this emission originates from two different sources, and the labels 12a-e and 13a-e were assigned based on the observed lensing configuration. To verify this interpretation and confirm that 12 and 13 are distinct objects, we additionally measured the rest-frame kinematic offsets between the two modes of Ly$\alpha$ emission red- and blue-shifted relative to the systemic redshift, similar to \citep{Hayes.Runnholm.ea2023}. These offsets were computed using \texttt{ppxf} \citep{Cappellari.Emsellem2004,Cappellari2017}, with the red- and blue-shifted peaks of Ly$\alpha$ emission modeled as two kinematically separate emission lines atop a 4th order polynomial background. The results of this analysis are shown in Figure \ref{fig:lya-kinematics}. In both stacked and individual spectra, Sources 12 and 13 appear kinematically distinct, with the blue Ly$\alpha$ emission in Source 12 offset significantly further from the system redshift.
In all four LAEs, red- and blue-shifted offsets range from $\sim 100$ to $\sim 400$ km/s, consistent with offsets at similar redshift \citep{McLinden.Finkelstein.ea2011}, as well as a larger study at lower redshift \citep{Hayes.Runnholm.ea2023}. As with Source 11, the interpretation of both Sources 12 and 13 as strongly lensed objects at the given redshifts is supported by the mass model in Paper II \cite{PaperII}.

For a point source emission line, the MUSE observations used in this work have a $5\sigma$ detection limit of roughly $10^{-18}$ erg s$^{-1}$ cm$^{-2}$, with mild wavelength dependence. All LAEs listed in this section exceed this limit, with all three images of the faintest source (Source 11) having a signal-to-noise ratio of roughly 10-20.

% SNR and limiting fluxes?
% JOD: I have a CSV table of fluxes and SNRs
% JOD: limiting flux for a 5sigma detection of a point source emission line is roughly 1e-18 erg s-1 cm-2, with mild wavelength dependence. 
% JOD: measured LyA fluxes range from ~1e-17 erg s-1 cm-2 (all Source 11 images), to ~2e-16 erg s-1 cm-2 (Source 12), using the same regions displayed in Fig 4. SNRs range from ~10 (Source 11) to >100 (Source 12)

\subsection{\label{sec:results_galaxies}Cluster Members and Field Galaxies}

\sisetup{
  round-mode=places,      % rounds numbers
  round-precision=3,      % keep 3 decimal places (change as needed)
}

\begin{table*}[h!]  % table 1, source redshifts for all objects in the MUSE and Gemini datasets with Q.O.P. > 3
\centering
\caption{Redshifts for all non-lensed objects in the MUSE and Gemini datasets with confident redshifts (Q.O.P. > 3). Object IDs correspond to Legacy Survey DR10 Brick \texttt{0910m360}. Spectroscopic analysis of object FGD suggests that it is a late-type emission galaxy.}
\setlength{\tabcolsep}{20pt} % default is ~6pt, smaller = table fills width more
\begin{filecontents*}{redshift_catalog_clean.txt}
770,0.49817,06:03:46.61,-35:57:29.08,4,GMOS
886,0.48663,06:03:47.36,-35:59:57.74,3,GMOS
1451,0.48663,06:03:50.89,-35:58:48.56,3,GMOS
2045,0.48715,06:03:54.45,-35:59:41.31,3,GMOS
2046,0.27137,06:03:54.45,-35:58:30.23,4,MUSE
2058,0.495,06:03:54.53,-35:57:46.15,4,MUSE
2067,0.95841,06:03:54.59,-35:57:56.26,4,MUSE
2136,0.49265,06:03:54.91,-35:57:38.99,3,GMOS
2188,0.4981,06:03:55.19,-35:58:26.59,4,MUSE
2234,0.49584,06:03:55.44,-35:58:05.55,4,MUSE
FGD,0.08607,06:03:55.56,-35:58:15.60,4,MUSE
2292,0.38454,06:03:55.77,-35:58:24.13,4,MUSE
2296,0.48824,06:03:55.78,-35:58:36.19,4,MUSE
2333,0.49399,06:03:55.92,-35:58:32.88,4,MUSE
2350,0.49046,06:03:55.99,-35:57:47.65,4,MUSE
2357,0.00006,06:03:56.03,-35:57:54.64,6,MUSE
2373,0.48785,06:03:56.13,-35:58:03.22,3,MUSE
2381,0.49971,06:03:56.17,-35:58:31.75,3,MUSE
2399,0.48731,06:03:56.28,-35:58:03.70,4,MUSE
2406,0.48507,06:03:56.32,-35:58:21.78,3,MUSE
2430,0.49656,06:03:56.42,-35:58:43.50,3,GMOS
2444,0.48829,06:03:56.51,-35:58:05.68,4,MUSE
2465,0.48444,06:03:56.60,-35:58:08.70,4,MUSE
2473,0.89784,06:03:56.67,-35:58:21.72,4,MUSE
2480,0.487,06:03:56.73,-35:58:34.03,4,MUSE
2521,0.48261,06:03:56.89,-35:58:15.30,4,MUSE
2533,0.49773,06:03:56.96,-35:58:10.10,3,MUSE
2542,1.4341,06:03:57.04,-35:58:39.78,4,MUSE
2543,0.48473,06:03:57.04,-35:58:15.43,4,MUSE
2546,0.48799,06:03:57.07,-35:59:18.28,4,GMOS
2573,0.49004,06:03:57.24,-35:58:08.69,4,MUSE
2588,0.48144,06:03:57.33,-35:57:59.22,4,MUSE
2593,0.49361,06:03:57.35,-35:58:05.90,3,MUSE
2605,0.48489,06:03:57.45,-35:57:20.17,3,GMOS
2617,0.49503,06:03:57.52,-35:57:57.06,3,MUSE
2667,0.49497,06:03:57.81,-35:58:14.85,3,MUSE
2677,0.47784,06:03:57.85,-35:58:30.98,4,MUSE
2710,0.48828,06:03:58.06,-35:58:13.89,3,MUSE
2747,0.48022,06:03:58.26,-35:58:14.54,4,MUSE
2788,0.47716,06:03:58.49,-35:58:27.06,4,MUSE
2823,0.4882,06:03:58.73,-35:58:13.34,3,MUSE
2830,0.49383,06:03:58.76,-35:58:34.47,4,MUSE
2873,0.48703,06:03:58.99,-35:57:41.15,4,MUSE
2881,0.49229,06:03:59.02,-35:58:20.06,4,MUSE
2946,0.95997,06:03:59.41,-35:57:58.03,3,MUSE
2951,0.49689,06:03:59.44,-35:57:59.96,3,MUSE
2959,0.48503,06:03:59.48,-35:56:21.61,4,GMOS
3025,0.49324,06:03:59.89,-35:57:22.13,3,GMOS
3111,0.4909,06:04:00.37,-35:55:43.37,3,GMOS
3196,0.49266,06:04:00.80,-35:57:43.98,3,GMOS
3424,0.49301,06:04:02.17,-35:57:14.62,3,GMOS
3582,0.49822,06:04:03.02,-35:58:30.08,3,GMOS
3617,0.491,06:04:03.24,-35:59:03.98,3,GMOS
3676,0.48584,06:04:03.52,-35:56:10.67,3,GMOS
3828,0.48656,06:04:04.51,-35:59:11.43,3,GMOS
4022,0.47817,06:04:05.62,-35:57:13.98,4,GMOS
4132,0.49144,06:04:06.31,-35:58:18.15,3,GMOS
4568,0.86463,06:04:08.96,-35:56:01.32,3,GMOS
\end{filecontents*}
\csvreader[
  separator=comma,
  head=false,
  tabular= c c c c c c,
  table head=\toprule Object ID & Redshift & RA (h:m:s) & Dec (d:m:s) & Instrument & Q.O.P. \\ \midrule,
  late after line=\\,
  table foot=\bottomrule
]{redshift_catalog_clean.txt}{}% 
{\csvcoli & \num{\csvcolii} & \csvcoliii & \csvcoliv & \csvcolvi & \csvcolv}
\label{tab:zcat}
\end{table*}

\begin{figure}[!h] % h = here, you can also use t, b, etc.
    \centering
    \includegraphics[width=0.9\columnwidth]{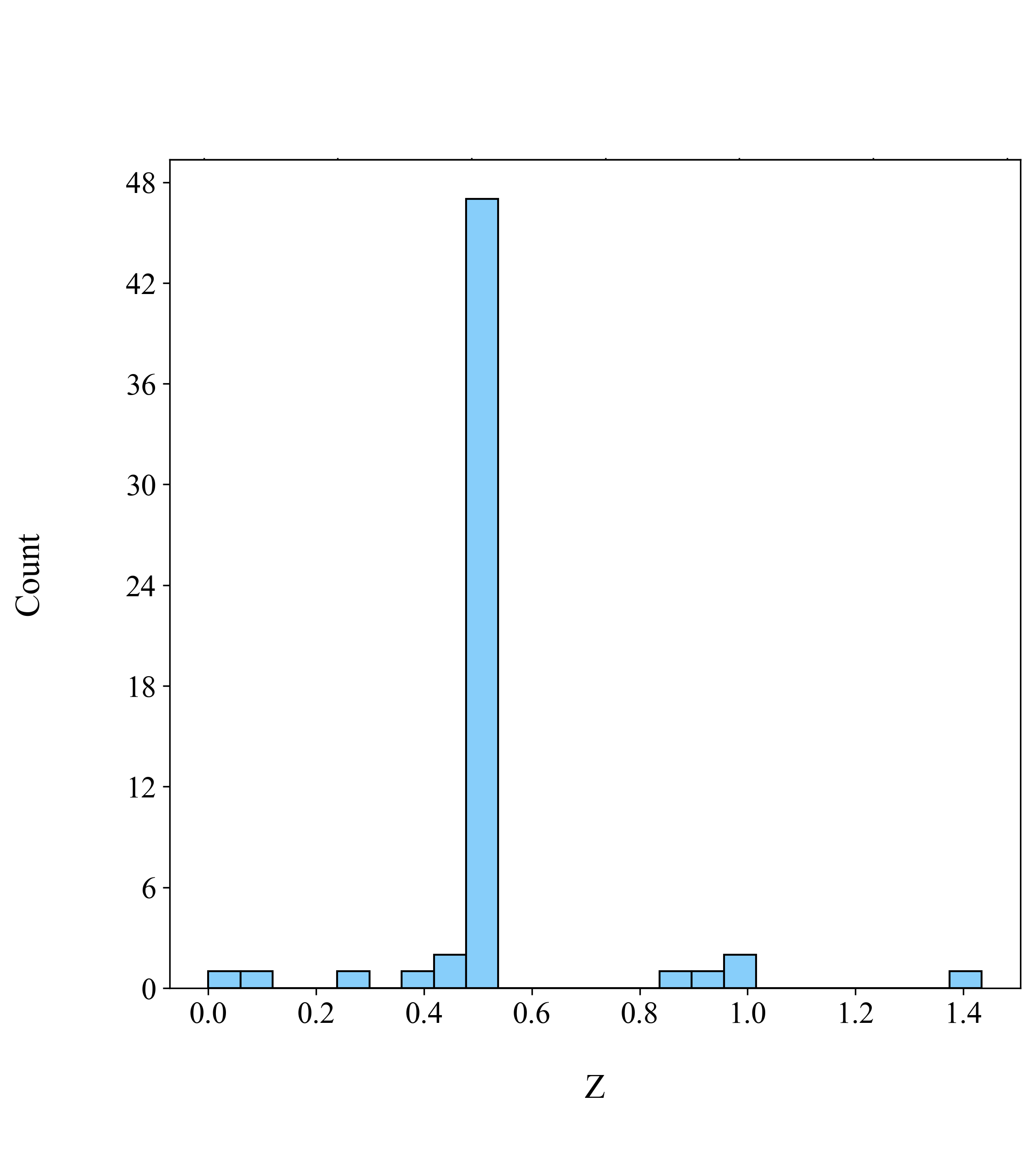}
    \includegraphics[width=0.9\columnwidth]{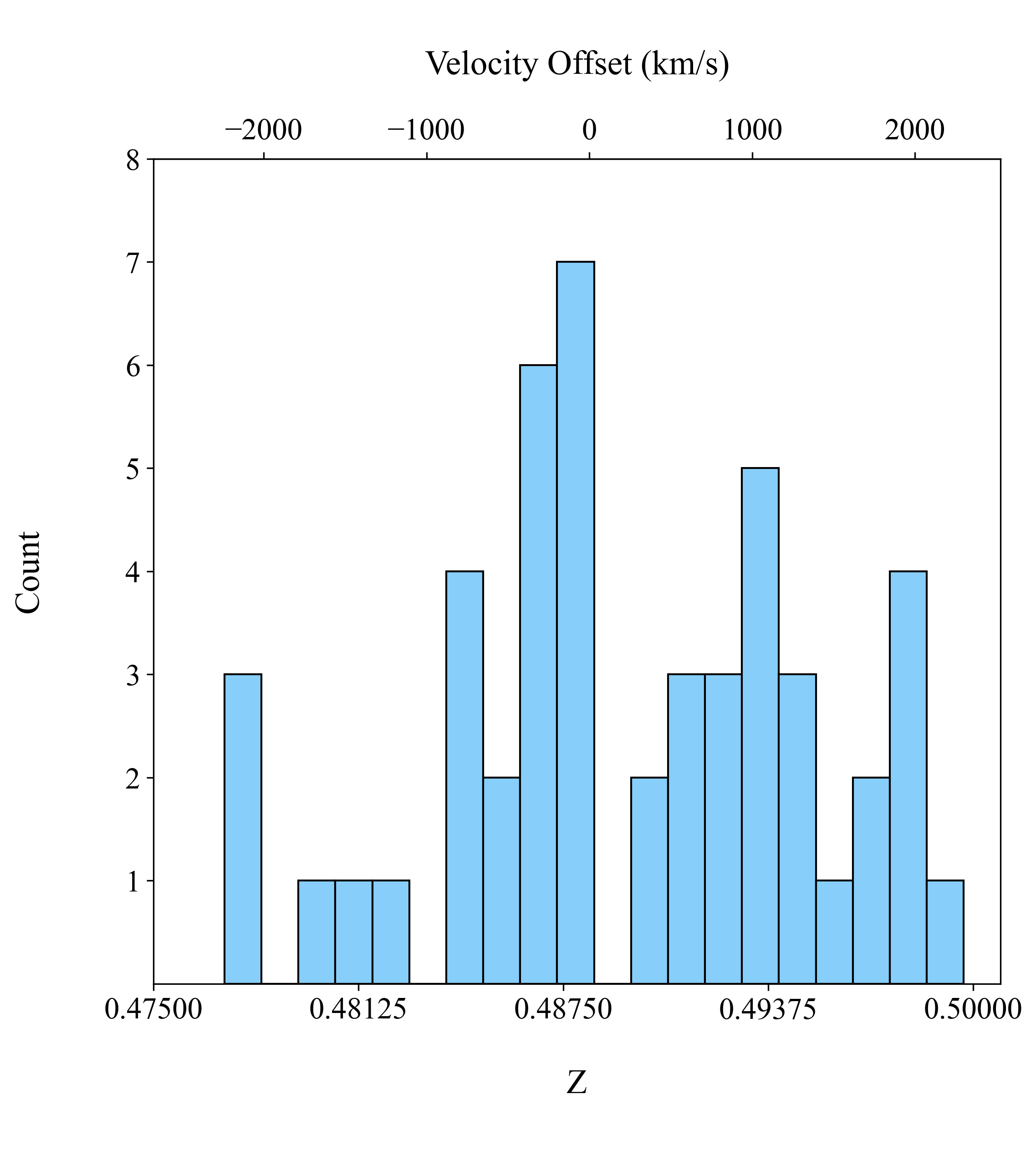}
    \caption{Histogram of confident redshifts for galaxies in the Carousel field. \reviewer{what do the authors mean with “Carousel field”? It would be interesting to see the full redshift distribution in the top panel, going up to the sources confirmed at z>4.} \textbf{Top}: All objects. \textbf{Bottom}: Likely cluster members, centered on the redshift of the Carousel cluster.}
    \label{fig:zhist}
\end{figure}

\begin{figure}[!h] % h = here, you can also use t, b, etc.
    \centering
    \includegraphics[width=\columnwidth]{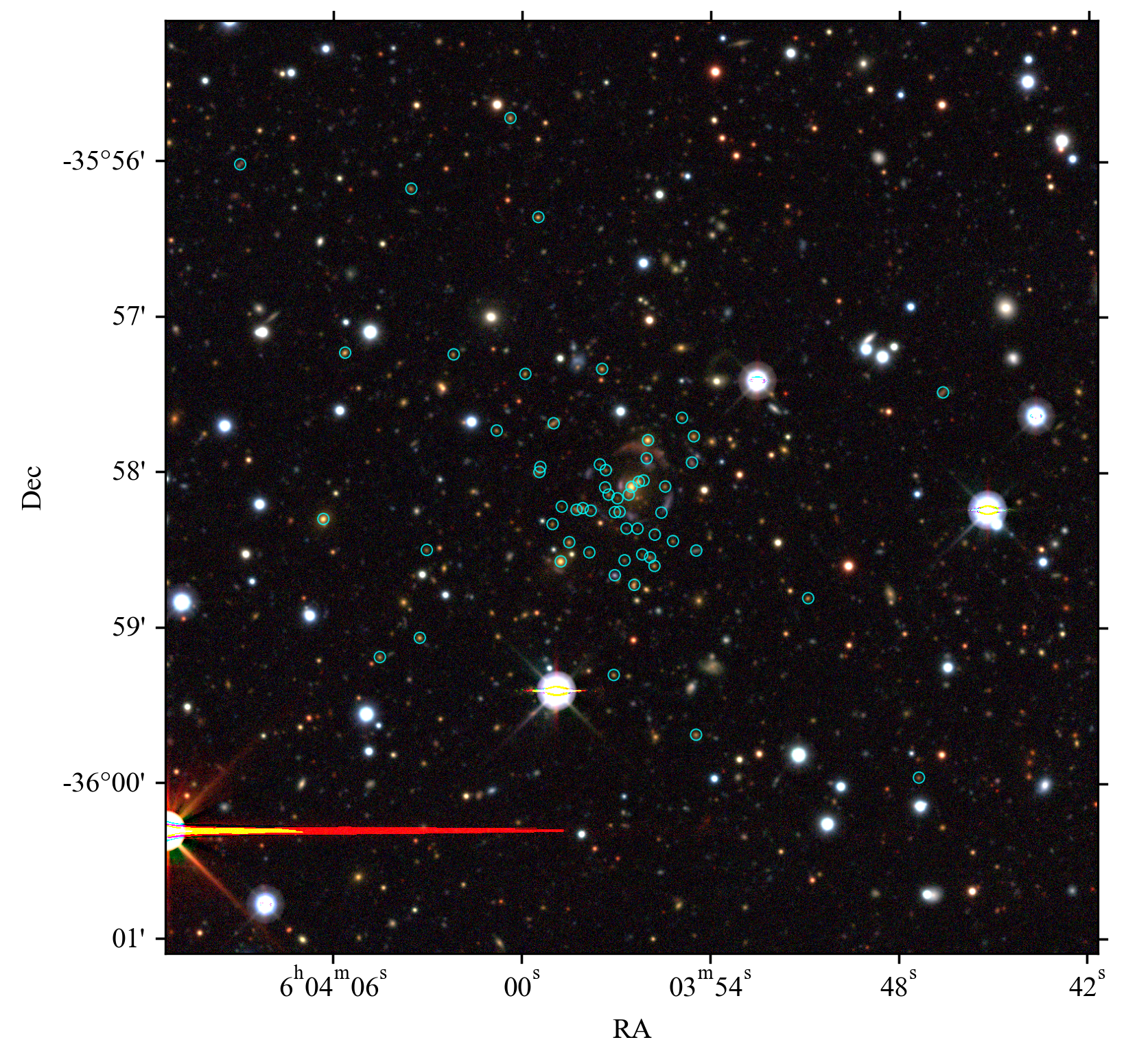}
    \includegraphics[width=\columnwidth]{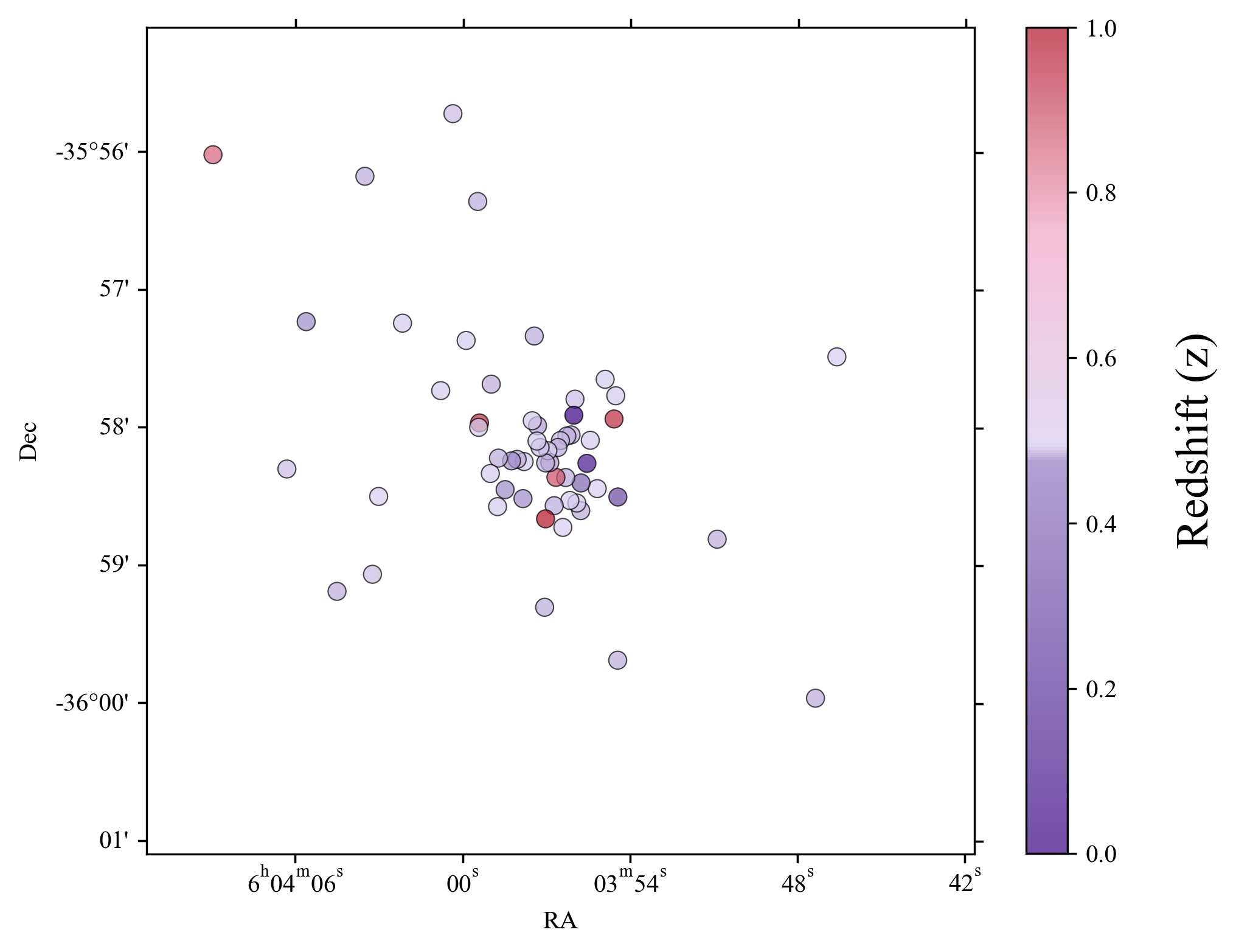}
    \caption{Sky map of all non-lensed galaxies in the MUSE and Gemini datasets with confident redshifts. \textbf{Top:} Catalog objects marked in cyan, overlaid on DECaLS imaging. \textbf{Bottom:} The same catalog and FOV, color coded by redshift.}
    \label{fig:skydist}
\end{figure}

We attempted to determine redshifts for all galaxies targeted by either MUSE or GMOS. 
All GMOS targets were matched to the DECaLS DR 10 object catalog \citep{LegacySurvey}, brick \texttt{0910m360}. %While the available HST imaging on this system is higher resolution, it is relatively shallow, and the DECaLS catalog is better suited to providing a set of target objects. \TODO{Jack: Need to verify previous claim somehow.} 
For all DECaLS objects within the MUSE field, we extract a spectrum by combining all spaxels within a 1'' radius of the object position. %\reviewer{ Is the chosen aperture circular with a radius of 1 arcsec? It is not clear from the text.} 
We exclude objects of type ``PSF'' as likely stars. All spectra were then examined using MARZ \citep{MARZ}; MARZ performs an automated template fitting for a set of galaxy and stellar templates of different types, and provides the top five redshift determinations.  These were examined independently by DYW and TEJ; in some cases, the redshifts were adjusted by hand to match prominent spectral features, particularly for galaxies at higher redshifts ($z \gtrsim 1$) where the template fitting became inaccurate. Redshifts were given a quality flag from 1-4, and only redshifts with quality 3 (good) and 4 (great) agreed on by both observers are reported here in Table \ref{tab:zcat}.  As the lensed sources are separately analyzed, DECaLS objects associated to these were removed from this table. Galaxies which were observed by both MUSE and Gemini were found to have consistent redshifts between the two, independent determinations (average offset of 0.0004), and where different the higher quality redshift is reported in Table \ref{tab:zcat}.  

In total, we determine redshifts for 57 DECaLS objects, of which one is a star and 56 are galaxies. 49 of these galaxies have redshifts consistent with the cluster redshift.  In addition, our analysis revealed a foreground dwarf galaxy at z=0.086 described below and listed as FGD in Table \ref{tab:zcat}.  The redshift distribution of galaxies in our catalog is shown in Figure \ref{fig:zhist}.

% Demetrius's work! Include breif mention of methods used to select objects

\subsubsection{\label{sec:results_dwarf}Foreground Dwarf Galaxy}

\begin{figure}[!h]
    \centering
    %\framebox(8cm,10cm){Source 4,6,FGD blend}
    %\includegraphics[width=0.9\linewidth]{figs/ICA_separation_46_blend.png}
    \includegraphics[width=0.9\linewidth]{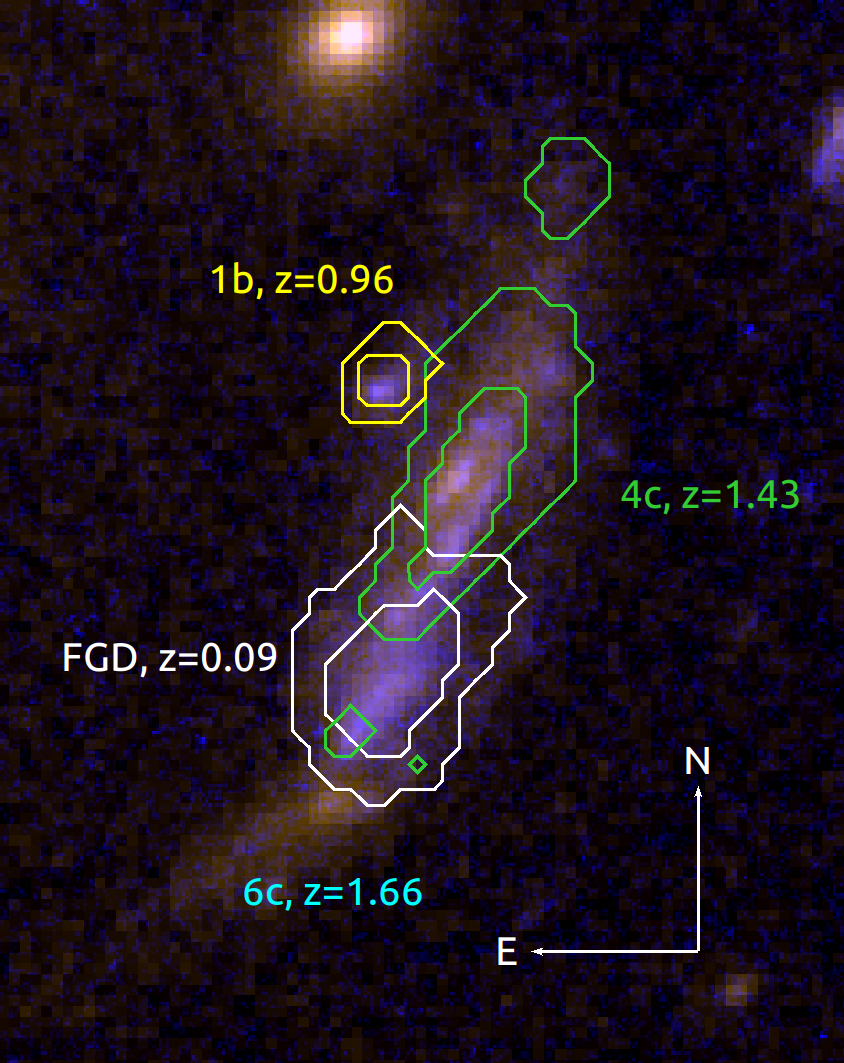}
    \caption{\textbf{Foreground Galaxy:} A region in the West of the Carousel field featuring three lensed sources in close proximity, highlighting the foreground emission-line galaxy blended with 4c and 6c. Contours show three sources as separated by ICA according to Section~\ref{sec:results_dwarf}, overlaid on HST imaging. The compass in the lower-right has arms of length 2''.}
    \label{fig:dwarf}
\end{figure}

As mentioned above, we identify a previously unknown foreground galaxy at z=0.086, blended with lensed images 4c and 6c. This galaxy was discovered when examining the spectrum of a DECaLS object associated to 4c.  In addition to the spectral features at the redshift of source 4, a set of additional emission features were found consistent with H$\alpha$, [OIII], and H$\beta$ at a redshift of 0.086. Examination revealed that the H$\alpha$ emission was prominent in a region between 4c and 6c where the color in the Hubble imaging is quite blue, and extraction of the spectrum in this region confirmed the spectral features at z=0.086.  Understanding the presence of this galaxy is essential to properly modeling the strong lensing of sources 4 and 6.

Additionally, in order to separate the spatial distribution of source images 1b, 4c, and the foreground galaxy, we applied an Independent Component Analysis (ICA) algorithm. A small rectangular region of the IFU data centered on this region was extracted, and the spectrum of each spaxel provided to the \texttt{FastICA} routine of \texttt{scikit-learn} \citep{scikit-learn}, allowing for 30 independent components. The results effectively separated the overlapping spatial distributions of sources 1b, 4c, and the foreground galaxy, contours of which are shown in Figure~\ref{fig:dwarf}. Source 6c was not detected by this process.

\subsection{\label{sec:results_vdisp}Cluster Velocity Dispersion}

% Demetrius's work! Mention 3000 km/s cuts. Describe bootstrap. Make double panel? add a median line and make lines thicker (median line should be different line style). Use formula in William Sheu paper to determine cluster mass.

We estimate the velocity dispersion of member galaxies within the cluster using both the biweight scale and gapper statistics from \cite{Beers90} and following the methodology in \cite{Wetzell22,Flowers2025}.  First, the biweight location is used to determine the central redshift of the cluster, which was found to be $z_{\text{cl}}=0.4895 \pm 0.0010$. The 49 galaxies within 3000 km s$^{-1}$ of the central redshift were then used to calculate the velocity dispersion; our procedure would call for removing galaxies more than three times the velocity dispersion from the central redshift and then recalculating the dispersion, but in practice this did not remove any additional galaxies.

Uncertainties on the velocity dispersion were estimated using a bootstrap procedure wherein the cluster member galaxies were resampled with replacement and the velocity dispersion was recalculated on this resampled data. This procedure was repeated 1,000 times, and the median and 68\% spread were taken as the velocity dispersion and its uncertainty.

\begin{figure}[!h] % h = here, you can also use t, b, etc.
    \centering
    \includegraphics[width=\columnwidth]{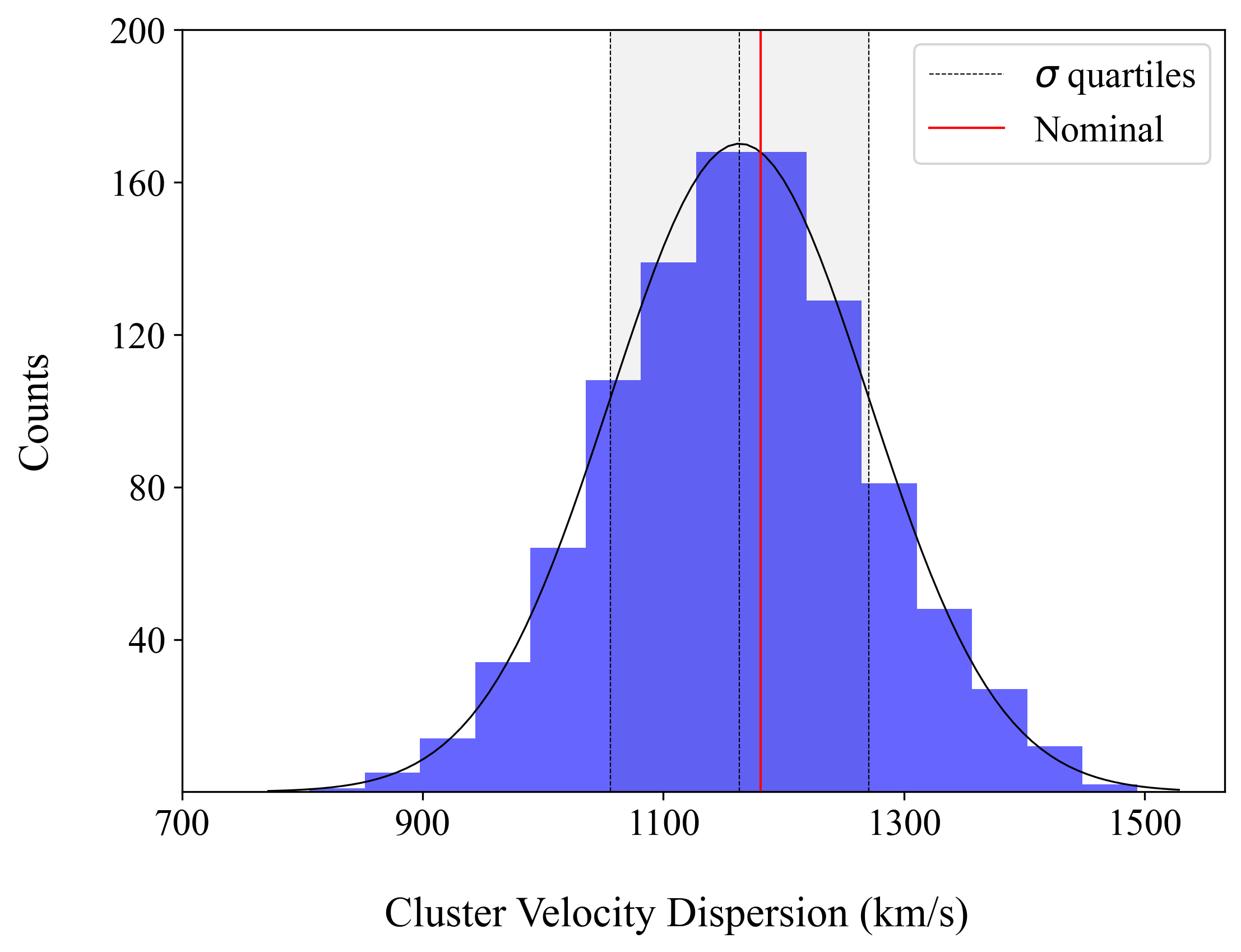}
    \includegraphics[width=\columnwidth]{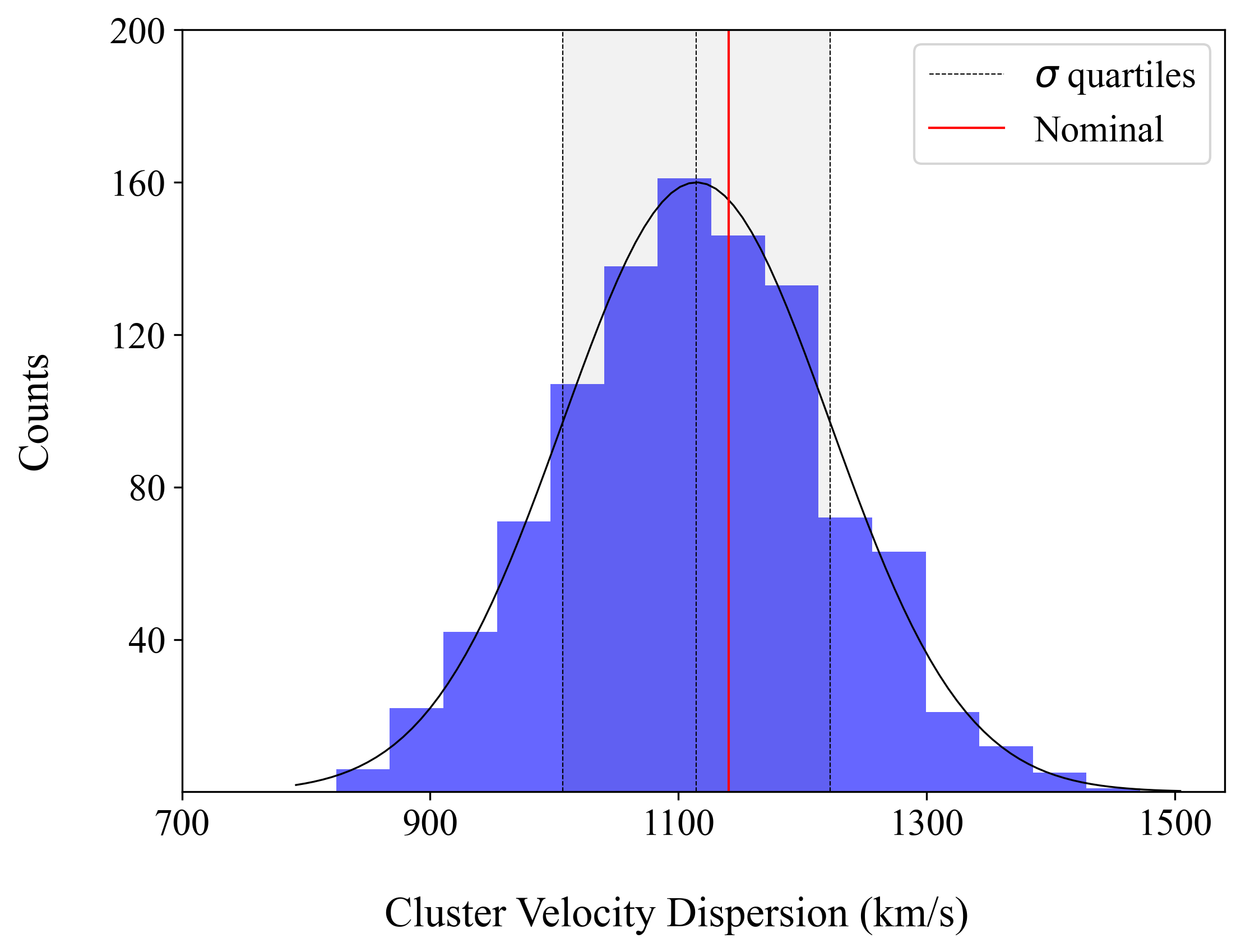}
    \caption{Top: Histogram of bootstrap results for the biweight method of determining the cluster velocity dispersion. Gray vertical lines and shaded region indicate the mean of the bootstrap results and the 68\% confidence interval; the red line shows the nominal result for the galaxy sample. Bottom: Same as the top panel, but for the gapper method of determining the cluster velocity dispersion.}
    \label{fig:sigma}
\end{figure}

%\begin{figure}[h] % h = here, you can also use t, b, etc.
%    \centering
 %   \includegraphics[width=\columnwidth]{figs/vel_disp_gap.png}
%    \caption{Histogram of bootstrap results for the gapper method of determining the cluster velocity dispersion, including vertical lines indicating the mean of the bootstrap results and the 68\% confidence interval.}
%    \label{Figure 6}
%\end{figure}

\begin{table}[!h]
\centering
\small
\renewcommand{\arraystretch}{2}
\setlength{\tabcolsep}{17pt}
\caption{Results for cluster velocity dispersion from the biweight and gapper methods. Column 2 gives the results for the nominal galaxy sample, while column 3 gives the median of the bootstrap trials and its respective uncertainty.}
\begin{tabular}{l c c}
\hline
\makecell{Dispersion \\ Method} & 
\makecell{Nominal \\ Result \\ (km s$^{-1}$)} & 
\makecell{Bootstrap \\ Median \\ (km s$^{-1}$)} \\
\hline
Biweight & 1181 & $1164^{+105}_{-106}$ \\
Gapper  & 1141 & $1114^{+108}_{-102}$ \\
\hline
\end{tabular}
\label{tab:sigma}
\end{table}

The results for both the biweight and gapper statistics are shown in Table \ref{tab:sigma}, and Figure \ref{fig:sigma} shows the distribution of velocity dispersions found in the bootstrap trials. Both the biweight and gapper statistics indicate a velocity dispersion of around 1100 km s$^{-1}$ with fully consistent results between methods and for the nominal sample compared to the bootstrap median.  The distributions of velocity dispersions for the bootstrap trials shown in Figure \ref{fig:sigma} are Gaussian with no indication of biases due to sample selection.

We use our estimate for the cluster velocity dispersion to estimate the $M_{200c}$ mass using the relation from \cite{Buckly-Geer11}. This relation is given by:

\begin{equation}
b_v^{1/\alpha} M_{200c} = 10^{15} M_\odot \, \frac{1}{h(z)} 
\left( \frac{\sigma_{\text{gal}}}{\sigma_{15}} \right)^{1/\alpha}
\end{equation}

\noindent where $b_{v}^{1/\alpha}$ is the unknown bias in the galaxy member velocities relative to the dark matter, $\sigma_{gal}$ is our estimated cluster velocity dispersion, the reference
velocity dispersion $\sigma_{15} = 1082.9 \pm 4$ km s$^{-1}$, and $\alpha=0.3361 \pm 0.0026$ is a dimensionless constant. %For $h(z)$, we assume a flat $\Lambda$CDM cosmology
We assume a flat $\Lambda$CDM cosmology with $\Omega_m = 0.3$ and $h = 0.69$.
%The reduced Hubble constant, $h(z)$, at redshift $z$ is $h(z) = h_o \sqrt{\Omega_{M} (1+z)^{3} + \Omega_{\Lambda}}$ where we take $\Omega_{\Lambda} = 0.7$, $\Omega_{M} = 0.3$, and $h_{o} = 0.69$.
Using the gapper median and uncertainties from our bootstrap trials for the velocity dispersion, we find $b_{v}^{1/\alpha} M_{200c} = 1.2_{-0.3}^{+0.4} \times 10^{15} M_{\odot}$.  This estimate of the mass is limited by the unknown bias between the galaxy and dark matter velocity dispersions which in turn depends on cluster mass, redshift, and the galaxy population used.  However, for massive halos simulations show this bias to be small; for example, \cite{Anbajagane22} find $b_{v}$ to be within 5\% of unity across different simulations, redshifts, and galaxy stellar mass cut for halos with $M_{200c} > 5\times 10^{14}$. This implies a bias in the mass of the Carousel cluster less than $\sim 15$\%.

\section{\label{sec:discussion}Discussion}

The Carousel lens is a uniquely powerful lens for constraints on cosmology, as demonstrated in upcoming Paper II in this series \cite{PaperII}.  Not only is it a relatively massive cluster at this redshift, but it is also a highly symmetric lens system leading to a relatively simple lens model compared to typical cluster lenses \citep{Sheu.Cikota.ea2024}.  It also displays a large number of lensed sources over a large redshift baseline with several different lens configurations, allowing for tight constraints on the mass model.

Our measured velocity cluster dispersion implies a very massive halo, around $10^{15} M_{\odot}$. This cluster was also detected in X-ray by the eROSITA All-Sky Survey (eRASS) \citep{Bulbul.Liu.ea2024}, and through the Sunyaev-Zel'dovich effect by the Atacama Cosmology Telescope (ACT) DR5 \citep{Klein.Mohr.ea2024}.
% The eRASS mass estimate is in the 92nd percentile M500 of that catalog
% And the 94th percentile M500 of the ACT catalog
Both eRASS and ACT-DR5 catalogs provide estimates of the cluster mass $M_{500c}$: the eRASS catalog estimate is $4.65^{+0.95}_{-0.73} \times 10^{14} M_{\odot}$, and the ACT-DR5 catalog provides an estimate of $5.59_{-0.91}^{+1.09} \times 10^{14} M_{\odot}$. Assuming the cosmological mass-concentration relation from \cite{Ludlow.Bose.ea2016a}, with a scatter on $\langle c_{500c}|M_{500c},z\rangle$ of 0.13 dex, the eRASS and ACT-DR5 mass estimates correspond to $M_{200c}$ of $6.87^{+1.51}_{-1.24} \times 10^{14} M_{\odot}$ and $8.28^{+1.84}_{-1.49} \times 10^{14} M_{\odot}$ respectively. Both X-ray and SZ mass estimates are somewhat lower than our kinematic estimate, though consistent. %\reviewer{A discussion on the systematic and statistical uncertainty related to this parameter $b_v^{1/\alpha}$ would benefit the discussion, as the mass is multiplied by this factor which scales $^(1/alpha)~3$. It could also alleviate the tension between the three estimates which, even if marginally consistent, differ of a factor ~2}.

The lens model derived in Paper 0 of this series gave an Einstein radius of $\theta_E=13.03$''$\pm0.02$'' for Source 4, at $z_4=1.432$. That work modeled the cluster halo mass as an elliptical power law, rather than an NFW, complicating comparison to the NFW mass estimates above. We therefore consider the projected aperture mass within $\theta_E$, which is $4.6 \times 10^{13} M_{\odot}$ for Source 4 using the same cosmology as above. A spherical halo of $M_{200c}=1.2 \times 10^{15}$ would reproduce this aperture mass with a concentration $c_{200c} \sim 10$, and lower mass halos would require even higher concentrations. While this is far above the expected cosmological mass-concentration relation, the selection effect of strong lensing is expected to preference highly concentrated halos. Additionally, baryonic mass in the inner halo, or elongation along the line-of-sight, would increase the projected mass in the inner halo and cause a higher apparent concentration.

Overall, the dynamical, lensing, X-ray, and SZ mass estimates are all consistent and indicate that the Carousel is a massive cluster.

%\newpage

\section{\label{sec:conclusion}Conclusion}

We have conducted a spectroscopic survey of the field surrounding the Carousel strong lensing cluster, using both Gemini/GMOS and VLT/MUSE, targeting both field galaxies and strongly lensed sources. We confirm the redshifts of the remaining sources identified in Paper 0 \citep{Sheu.Cikota.ea2024}, and identify 6 new lensed sources, bringing the total to 12.
Of these 12, only one (Source 10) remains without redshift. The sources with known redshifts range from $z=0.96$ to $z=4.1$ and nearly all of them display at least three images. Sources 4, 12, and 13 have five identified images including a central image, and sources 6 and 7 are quads. The full list of sources and their redshifts are given in the Table~\ref{tab:full-source-list}, along with spectra of sources with newly reported redshifts.

We identify four strongly lensed Lyman $\alpha$ emitters, three of which are not apparent in imaging data and are only detected in MUSE IFU data. These sources span the redshift range $3.09 < z < 4.09$, and two of them exhibit clear central images near the BCG of the Carousel cluster. These sources could be of considerable interest for cosmology constraints in the Carousel: their central images could strongly constrain the dark matter density profile of the cluster \citep[as in e.g.][]{Manjon-Garcia.Diego.ea2020a,Cerny.Jauzac.ea2025}, and their high redshift will allow multi-source-plane cosmography studies to constrain the universe's expansion history at much earlier times. Future deep near infrared imaging may precisely reveal these sources' positions and morphology, enabling higher precision strong lens models.

In addition, we identify 49 cluster members as well as additional field galaxies.  We find a cluster velocity dispersion of $1100\pm100$ km s$^{-1}$ giving an estimated mass of $b_{v}^{1/a} M_{200c} = 1.2_{-0.3}^{+0.4} \times 10^{15} M_{\odot}$. This is consistent with, though somewhat higher than, mass estimates from both the eRASS X-ray and ACT-DR5 SZ catalogs, which give $M_{200c} = 6.9^{+1.5}_{-1.2}\times 10^{14} M_{\odot}$ and $8.3^{+1.8}_{-1.5} \times 10^{14} M_{\odot}$ respectively.

Spectroscopic follow-up of the Carousel system has so far spanned the wavelengths
% MUSE: ~470nm to ~935nm
% GMOS: varies, typically ~520nm to 1.02um
$\sim 470$nm to $\sim 1\mu$m. This wavelength range is sensitive to the common emission lines [O II] 3729\AA, [O III] 5007\AA, or H$\alpha$ at $z\lesssim1.7$, and Ly$\alpha$ at $2.9\lesssim z\lesssim 6.7$. Further spectroscopic observations at other wavelengths will potentially identify additional lensed sources in the `redshift desert' $1.7 < z < 2.9$, or at very high redshift $z \gtrsim 6.7$.

\begin{acknowledgments}
This work made extensive use of open-source software libraries. 
In addition to those noted throughout the text, these include \texttt{astropy} \citep{astropy-01,astropy-02,astropy-03}, \texttt{numpy} \citep{numpy}, \texttt{scipy} \citep{scipy}, and \texttt{matplotlib} \citep{matplotlib}.
We acknowledge use of the lux supercomputer at UC Santa Cruz, funded by NSF MRI grant AST 1828315.
\end{acknowledgments}

%\appendix

%\section{\label{app:all-sources}Full Source Catalog and Spectra}

%Here we present a full catalog of lensed sources in the field, including those previously reported in Paper 0 \citep{Sheu.Cikota.ea2024}. Table~\ref{tab:full-source-list} gives the name, position, and redshift of each strongly-lensed source, as well as spectral features from which the redshifts were derived. Note that coordinates of each source are provided for clarity, and are not intended as precision measurements.

%We additionally present the MUSE and GMOS spectra of lensed sources with newly published redshifts. Sources previously identified from imaging data are shown in Figure~\ref{fig:source-spectra}, and the four LAEs are presented in Figure~\ref{fig:LAE-spectra}.

\bibliographystyle{apsrev4-1}

\bibliography{carousel}

@article{Sheu.Cikota.ea2024,
  title = {The {{Carousel Lens}}: {{A Well-modeled Strong Lens}} with {{Multiple Sources Spectroscopically Confirmed}} by {{VLT}}/{{MUSE}}},
  shorttitle = {The {{Carousel Lens}}},
  author = {Sheu, William and Cikota, Aleksandar and Huang, Xiaosheng and Glazebrook, Karl and Storfer, Christopher and Agarwal, Shrihan and Schlegel, David J. and Suzuki, Nao and Barone, Tania M. and Bian, Fuyan and Jeltema, Tesla and Jones, Tucker and Kacprzak, Glenn G. and O'Donnell, Jackson H. and G. C., Keerthi Vasan},
  year = {2024},
  month = sep,
  journal = {The Astrophysical Journal},
  volume = {973},
  pages = {3},
  issn = {0004-637X},
  doi = {10.3847/1538-4357/ad65d3},
  urldate = {2025-09-08}
}

@article{MARZ,
  title = {{{MARZ}}: {{Manual}} and Automatic Redshifting Software},
  shorttitle = {{{MARZ}}},
  author = {Hinton, S. R. and Davis, Tamara M. and Lidman, C. and Glazebrook, K. and Lewis, G. F.},
  year = {2016},
  month = apr,
  journal = {Astronomy and Computing},
  volume = {15},
  pages = {61--71},
  issn = {2213-1337},
  doi = {10.1016/j.ascom.2016.03.001},
  urldate = {2024-08-15}
}

@article{PypeIt_JOSS,
  title = {{{PypeIt}}: {{The Python Spectroscopic Data Reduction Pipeline}}},
  shorttitle = {{{PypeIt}}},
  author = {Prochaska, J. and Hennawi, Joseph and Westfall, Kyle and Cooke, Ryan and Wang, Feige and Hsyu, Tiffany and Davies, Frederick and Farina, Emanuele and Pelliccia, Debora},
  year = {2020},
  month = dec,
  journal = {The Journal of Open Source Software},
  volume = {5},
  pages = {2308},
  doi = {10.21105/joss.02308},
  urldate = {2022-12-03}
}

@software{PypeIt_Zenodo,
       author = {{Prochaska}, J. Xavier and {Hennawi}, Joseph and {Cooke}, Ryan and {Westfall}, Kyle and {Wang}, Feige and {EmAstro} and {Tiffanyhsyu} and {Wasserman}, Asher and {Villaume}, Alexa and {Marijana777} and {Schindler}, JT and {Young}, David and {Simha}, Sunil and {Wilde}, Matt and {Tejos}, Nicolas and {Isbell}, Jacob and {Fl{\"o}rs}, Andreas and {Sandford}, Nathan and {Vasovi{\'c}}, Zlatan and {Betts}, Edward and {Holden}, Brad},
        title = "{pypeit/PypeIt: Release 1.0.0}",
         year = {2020},
        month = apr,
          eid = {10.5281/zenodo.3743493},
          doi = {10.5281/zenodo.3743493},
      version = {v1.0.0},
    publisher = {Zenodo},
       adsurl = {https://ui.adsabs.harvard.edu/abs/2020zndo...3743493P}
}

@ARTICLE{Collett2012,
       author = {{Collett}, T.~E. and {Auger}, M.~W. and {Belokurov}, V. and {Marshall}, P.~J. and {Hall}, A.~C.},
        title = "{Constraining the dark energy equation of state with double-source plane strong lenses}",
      journal = {\mnras},
         year = 2012,
        month = aug,
       volume = {424},
       number = {4},
        pages = {2864-2875},
          doi = {10.1111/j.1365-2966.2012.21424.x},
archivePrefix = {arXiv},
       eprint = {1203.2758},
 primaryClass = {astro-ph.CO},
       adsurl = {https://ui.adsabs.harvard.edu/abs/2012MNRAS.424.2864C}
}

@ARTICLE{Caminha2022,
       author = {{Caminha}, G.~B. and {Suyu}, S.~H. and {Grillo}, C. and {Rosati}, P.},
        title = "{Galaxy cluster strong lensing cosmography. Cosmological constraints from a sample of regular galaxy clusters}",
      journal = {\aap},
         year = 2022,
        month = jan,
       volume = {657},
          eid = {A83},
        pages = {A83},
          doi = {10.1051/0004-6361/202141994},
archivePrefix = {arXiv},
       eprint = {2110.06232},
 primaryClass = {astro-ph.CO},
       adsurl = {https://ui.adsabs.harvard.edu/abs/2022A&A...657A..83C}
}

@ARTICLE{Jullo2010,
       author = {{Jullo}, E. and {Natarajan}, P. and {Kneib}, J. -P. and {D'Aloisio}, A. and {Limousin}, M. and {Richard}, J. and {Schimd}, C.},
        title = "{Cosmological constraints from strong gravitational lensing in clusters of galaxies.}",
      journal = {Science},
         year = 2010,
        month = aug,
       volume = {329},
        pages = {924-927},
          doi = {10.1126/science.1185759},
archivePrefix = {arXiv},
       eprint = {1008.4802},
 primaryClass = {astro-ph.CO},
       adsurl = {https://ui.adsabs.harvard.edu/abs/2010Sci...329..924J}
}

@ARTICLE{Bolamperti2024,
       author = {{Bolamperti}, A. and {Grillo}, C. and {Caminha}, G.~B. and {Granata}, G. and {Suyu}, S.~H. and {Ca{\~n}ameras}, R. and {Christensen}, L. and {Vernet}, J. and {Zanella}, A.},
        title = "{Cosmography from accurate mass modeling of the lens group SDSS J0100+1818: Five sources at three different redshifts}",
      journal = {\aap},
         year = 2024,
        month = dec,
       volume = {692},
          eid = {A239},
        pages = {A239},
          doi = {10.1051/0004-6361/202451209},
archivePrefix = {arXiv},
       eprint = {2411.07289},
 primaryClass = {astro-ph.CO},
       adsurl = {https://ui.adsabs.harvard.edu/abs/2024A&A...692A.239B}
}

@ARTICLE{Caminha2016,
       author = {{Caminha}, G.~B. and {Grillo}, C. and {Rosati}, P. and {Balestra}, I. and {Karman}, W. and {Lombardi}, M. and {Mercurio}, A. and {Nonino}, M. and {Tozzi}, P. and {Zitrin}, A. and {Biviano}, A. and {Girardi}, M. and {Koekemoer}, A.~M. and {Melchior}, P. and {Meneghetti}, M. and {Munari}, E. and {Suyu}, S.~H. and {Umetsu}, K. and {Annunziatella}, M. and {Borgani}, S. and {Broadhurst}, T. and {Caputi}, K.~I. and {Coe}, D. and {Delgado-Correal}, C. and {Ettori}, S. and {Fritz}, A. and {Frye}, B. and {Gobat}, R. and {Maier}, C. and {Monna}, A. and {Postman}, M. and {Sartoris}, B. and {Seitz}, S. and {Vanzella}, E. and {Ziegler}, B.},
        title = "{CLASH-VLT: A highly precise strong lensing model of the galaxy cluster RXC J2248.7-4431 (Abell S1063) and prospects for cosmography}",
      journal = {\aap},
         year = 2016,
        month = mar,
       volume = {587},
          eid = {A80},
        pages = {A80},
          doi = {10.1051/0004-6361/201527670},
archivePrefix = {arXiv},
       eprint = {1512.04555},
 primaryClass = {astro-ph.CO},
       adsurl = {https://ui.adsabs.harvard.edu/abs/2016A&A...587A..80C}
}

@ARTICLE{Huang2021,
       author = {{Huang}, X. and {Storfer}, C. and {Gu}, A. and {Ravi}, V. and {Pilon}, A. and {Sheu}, W. and {Venguswamy}, R. and {Banka}, S. and {Dey}, A. and {Landriau}, M. and {Lang}, D. and {Meisner}, A. and {Moustakas}, J. and {Myers}, A.~D. and {Sajith}, R. and {Schlafly}, E.~F. and {Schlegel}, D.~J.},
        title = "{Discovering New Strong Gravitational Lenses in the DESI Legacy Imaging Surveys}",
      journal = {\apj},
         year = 2021,
        month = mar,
       volume = {909},
       number = {1},
          eid = {27},
        pages = {27},
          doi = {10.3847/1538-4357/abd62b},
archivePrefix = {arXiv},
       eprint = {2005.04730},
 primaryClass = {astro-ph.IM},
       adsurl = {https://ui.adsabs.harvard.edu/abs/2021ApJ...909...27H}
}

@ARTICLE{Bowden2025,
       author = {{Bowden}, Duncan J. and {Sahu}, Nandini and {Shajib}, Anowar J. and {Tran}, Kim-Vy and {Barone}, Tania M. and {Keerthi Vasan G.}, C. and {Ballard}, Daniel J. and {Collett}, Thomas E. and {Dalessandro}, Faith and {Ferrami}, Giovanni and {Glazebrook}, Karl and {Gottemoller}, William J. and {Iwamoto}, Leena and {Jones}, Tucker and {Kacprzak}, Glenn G. and {Lewis}, Geraint F. and {McIntosh-Lombardo}, Haven and {Skobe}, Hannah and {Suyu}, Sherry H. and {Sweet}, Sarah M.},
        title = "{Constraining Cosmology with Double-Source-Plane Strong Gravitational Lenses From the AGEL Survey}",
      journal = {arXiv e-prints},
         year = 2025,
        month = sep,
          eid = {arXiv:2509.15012},
        pages = {arXiv:2509.15012},
          doi = {10.48550/arXiv.2509.15012},
archivePrefix = {arXiv},
       eprint = {2509.15012},
 primaryClass = {astro-ph.CO},
       adsurl = {https://ui.adsabs.harvard.edu/abs/2025arXiv250915012B}
}

@ARTICLE{Sahu2025,
       author = {{Sahu}, Nandini and {Shajib}, Anowar J. and {Tran}, Kim-Vy and {Skobe}, Hannah and {Rhoades}, Sunny and {Jones}, Tucker and {Glazebrook}, Karl and {Collett}, Thomas E. and {Suyu}, Sherry H. and {Barone}, Tania M. and {Keerthi Vasan G.}, C. and {Bowden}, Duncan J. and {Ballard}, Daniel and {Kacprzak}, Glenn G. and {Sweet}, Sarah M. and {Lewis}, Geraint F. and {Nanayakkara}, Themiya},
        title = "{Cosmography with the Double Source Plane Strong Gravitational Lens AGEL150745+052256}",
      journal = {arXiv e-prints},
         year = 2025,
        month = apr,
          eid = {arXiv:2504.00656},
        pages = {arXiv:2504.00656},
          doi = {10.48550/arXiv.2504.00656},
archivePrefix = {arXiv},
       eprint = {2504.00656},
 primaryClass = {astro-ph.CO},
       adsurl = {https://ui.adsabs.harvard.edu/abs/2025arXiv250400656S}
}

@ARTICLE{Collett2014,
       author = {{Collett}, Thomas E. and {Auger}, Matthew W.},
        title = "{Cosmological constraints from the double source plane lens SDSSJ0946+1006}",
      journal = {\mnras},
         year = 2014,
        month = sep,
       volume = {443},
       number = {2},
        pages = {969-976},
          doi = {10.1093/mnras/stu1190},
archivePrefix = {arXiv},
       eprint = {1403.5278},
 primaryClass = {astro-ph.CO},
       adsurl = {https://ui.adsabs.harvard.edu/abs/2014MNRAS.443..969C}
}

@ARTICLE{MUSEZAP,
       author = {{Soto}, Kurt T. and {Lilly}, Simon J. and {Bacon}, Roland and {Richard}, Johan and {Conseil}, Simon},
        title = "{ZAP - enhanced PCA sky subtraction for integral field spectroscopy}",
      journal = {\mnras},
         year = 2016,
        month = may,
       volume = {458},
       number = {3},
        pages = {3210-3220},
          doi = {10.1093/mnras/stw474},
archivePrefix = {arXiv},
       eprint = {1602.08037},
 primaryClass = {astro-ph.IM},
       adsurl = {https://ui.adsabs.harvard.edu/abs/2016MNRAS.458.3210S}
}

@article{LegacySurvey,
  title = {Overview of the {{DESI Legacy Imaging Surveys}}},
  author = {Dey, Arjun and Schlegel, David J. and Lang, Dustin and Blum, Robert and Burleigh, Kaylan and Fan, Xiaohui and Findlay, Joseph R. and Finkbeiner, Doug and Herrera, David and Juneau, St{\'e}phanie and Landriau, Martin and Levi, Michael and McGreer, Ian and Meisner, Aaron and Myers, Adam D. and Moustakas, John and Nugent, Peter and Patej, Anna and Schlafly, Edward F. and Walker, Alistair R. and Valdes, Francisco and Weaver, Benjamin A. and Y{\`e}che, Christophe and Zou, Hu and Zhou, Xu and Abareshi, Behzad and Abbott, T. M. C. and Abolfathi, Bela and Aguilera, C. and Alam, Shadab and Allen, Lori and Alvarez, A. and Annis, James and Ansarinejad, Behzad and Aubert, Marie and Beechert, Jacqueline and Bell, Eric F. and BenZvi, Segev Y. and Beutler, Florian and Bielby, Richard M. and Bolton, Adam S. and Brice{\~n}o, C{\'e}sar and {Buckley-Geer}, Elizabeth J. and Butler, Karen and Calamida, Annalisa and Carlberg, Raymond G. and Carter, Paul and Casas, Ricard and Castander, Francisco J. and Choi, Yumi and Comparat, Johan and Cukanovaite, Elena and Delubac, Timoth{\'e}e and DeVries, Kaitlin and Dey, Sharmila and Dhungana, Govinda and Dickinson, Mark and Ding, Zhejie and Donaldson, John B. and Duan, Yutong and Duckworth, Christopher J. and Eftekharzadeh, Sarah and Eisenstein, Daniel J. and Etourneau, Thomas and Fagrelius, Parker A. and Farihi, Jay and Fitzpatrick, Mike and {Font-Ribera}, Andreu and Fulmer, Leah and G{\"a}nsicke, Boris T. and Gaztanaga, Enrique and George, Koshy and Gerdes, David W. and Gontcho, Satya Gontcho A. and Gorgoni, Claudio and Green, Gregory and Guy, Julien and Harmer, Diane and Hernandez, M. and Honscheid, Klaus and Huang, Lijuan Wendy and James, David J. and Jannuzi, Buell T. and Jiang, Linhua and Joyce, Richard and Karcher, Armin and Karkar, Sonia and Kehoe, Robert and Kneib, Jean-Paul and {Kueter-Young}, Andrea and Lan, Ting-Wen and Lauer, Tod R. and Le Guillou, Laurent and Le Van Suu, Auguste and Lee, Jae Hyeon and Lesser, Michael and Perreault Levasseur, Laurence and Li, Ting S. and Mann, Justin L. and Marshall, Robert and {Mart{\'i}nez-V{\'a}zquez}, C. E. and Martini, Paul and {du Mas des Bourboux}, H{\'e}lion and McManus, Sean and Meier, Tobias Gabriel and M{\'e}nard, Brice and Metcalfe, Nigel and {Mu{\~n}oz-Guti{\'e}rrez}, Andrea and Najita, Joan and Napier, Kevin and Narayan, Gautham and Newman, Jeffrey A. and Nie, Jundan and Nord, Brian and Norman, Dara J. and Olsen, Knut A. G. and Paat, Anthony and {Palanque-Delabrouille}, Nathalie and Peng, Xiyan and Poppett, Claire L. and Poremba, Megan R. and Prakash, Abhishek and Rabinowitz, David and Raichoor, Anand and Rezaie, Mehdi and Robertson, A. N. and Roe, Natalie A. and Ross, Ashley J. and Ross, Nicholas P. and Rudnick, Gregory and Safonova, Sasha and Saha, Abhijit and S{\'a}nchez, F. Javier and Savary, Elodie and Schweiker, Heidi and Scott, Adam and Seo, Hee-Jong and Shan, Huanyuan and Silva, David R. and Slepian, Zachary and Soto, Christian and Sprayberry, David and Staten, Ryan and Stillman, Coley M. and Stupak, Robert J. and Summers, David L. and Sien Tie, Suk and Tirado, H. and {Vargas-Maga{\~n}a}, Mariana and Vivas, A. Katherina and Wechsler, Risa H. and Williams, Doug and Yang, Jinyi and Yang, Qian and Yapici, Tolga and Zaritsky, Dennis and Zenteno, A. and Zhang, Kai and Zhang, Tianmeng and Zhou, Rongpu and Zhou, Zhimin},
  year = {2019},
  month = may,
  journal = {The Astronomical Journal},
  volume = {157},
  pages = {168},
  publisher = {IOP},
  issn = {0004-6256},
  doi = {10.3847/1538-3881/ab089d},
  urldate = {2025-07-25}
}

@ARTICLE{GAIA-DR3,
       author = {{Gaia Collaboration} and {Vallenari}, A. and {Brown}, A.~G.~A. and {Prusti}, T. and {de Bruijne}, J.~H.~J. and {Arenou}, F. and {Babusiaux}, C. and {Biermann}, M. and {Creevey}, O.~L. and {Ducourant}, C. and {Evans}, D.~W. and {Eyer}, L. and {Guerra}, R. and {Hutton}, A. and {Jordi}, C. and {Klioner}, S.~A. and {Lammers}, U.~L. and {Lindegren}, L. and {Luri}, X. and {Mignard}, F. and {Panem}, C. and {Pourbaix}, D. and {Randich}, S. and {Sartoretti}, P. and {Soubiran}, C. and {Tanga}, P. and {Walton}, N.~A. and {Bailer-Jones}, C.~A.~L. and {Bastian}, U. and {Drimmel}, R. and {Jansen}, F. and {Katz}, D. and {Lattanzi}, M.~G. and {van Leeuwen}, F. and {Bakker}, J. and {Cacciari}, C. and {Casta{\~n}eda}, J. and {De Angeli}, F. and {Fabricius}, C. and {Fouesneau}, M. and {Fr{\'e}mat}, Y. and {Galluccio}, L. and {Guerrier}, A. and {Heiter}, U. and {Masana}, E. and {Messineo}, R. and {Mowlavi}, N. and {Nicolas}, C. and {Nienartowicz}, K. and {Pailler}, F. and {Panuzzo}, P. and {Riclet}, F. and {Roux}, W. and {Seabroke}, G.~M. and {Sordo}, R. and {Th{\'e}venin}, F. and {Gracia-Abril}, G. and {Portell}, J. and {Teyssier}, D. and {Altmann}, M. and {Andrae}, R. and {Audard}, M. and {Bellas-Velidis}, I. and {Benson}, K. and {Berthier}, J. and {Blomme}, R. and {Burgess}, P.~W. and {Busonero}, D. and {Busso}, G. and {C{\'a}novas}, H. and {Carry}, B. and {Cellino}, A. and {Cheek}, N. and {Clementini}, G. and {Damerdji}, Y. and {Davidson}, M. and {de Teodoro}, P. and {Nu{\~n}ez Campos}, M. and {Delchambre}, L. and {Dell'Oro}, A. and {Esquej}, P. and {Fern{\'a}ndez-Hern{\'a}ndez}, J. and {Fraile}, E. and {Garabato}, D. and {Garc{\'\i}a-Lario}, P. and {Gosset}, E. and {Haigron}, R. and {Halbwachs}, J. -L. and {Hambly}, N.~C. and {Harrison}, D.~L. and {Hern{\'a}ndez}, J. and {Hestroffer}, D. and {Hodgkin}, S.~T. and {Holl}, B. and {Jan{\ss}en}, K. and {Jevardat de Fombelle}, G. and {Jordan}, S. and {Krone-Martins}, A. and {Lanzafame}, A.~C. and {L{\"o}ffler}, W. and {Marchal}, O. and {Marrese}, P.~M. and {Moitinho}, A. and {Muinonen}, K. and {Osborne}, P. and {Pancino}, E. and {Pauwels}, T. and {Recio-Blanco}, A. and {Reyl{\'e}}, C. and {Riello}, M. and {Rimoldini}, L. and {Roegiers}, T. and {Rybizki}, J. and {Sarro}, L.~M. and {Siopis}, C. and {Smith}, M. and {Sozzetti}, A. and {Utrilla}, E. and {van Leeuwen}, M. and {Abbas}, U. and {{\'A}brah{\'a}m}, P. and {Abreu Aramburu}, A. and {Aerts}, C. and {Aguado}, J.~J. and {Ajaj}, M. and {Aldea-Montero}, F. and {Altavilla}, G. and {{\'A}lvarez}, M.~A. and {Alves}, J. and {Anders}, F. and {Anderson}, R.~I. and {Anglada Varela}, E. and {Antoja}, T. and {Baines}, D. and {Baker}, S.~G. and {Balaguer-N{\'u}{\~n}ez}, L. and {Balbinot}, E. and {Balog}, Z. and {Barache}, C. and {Barbato}, D. and {Barros}, M. and {Barstow}, M.~A. and {Bartolom{\'e}}, S. and {Bassilana}, J. -L. and {Bauchet}, N. and {Becciani}, U. and {Bellazzini}, M. and {Berihuete}, A. and {Bernet}, M. and {Bertone}, S. and {Bianchi}, L. and {Binnenfeld}, A. and {Blanco-Cuaresma}, S. and {Blazere}, A. and {Boch}, T. and {Bombrun}, A. and {Bossini}, D. and {Bouquillon}, S. and {Bragaglia}, A. and {Bramante}, L. and {Breedt}, E. and {Bressan}, A. and {Brouillet}, N. and {Brugaletta}, E. and {Bucciarelli}, B. and {Burlacu}, A. and {Butkevich}, A.~G. and {Buzzi}, R. and {Caffau}, E. and {Cancelliere}, R. and {Cantat-Gaudin}, T. and {Carballo}, R. and {Carlucci}, T. and {Carnerero}, M.~I. and {Carrasco}, J.~M. and {Casamiquela}, L. and {Castellani}, M. and {Castro-Ginard}, A. and {Chaoul}, L. and {Charlot}, P. and {Chemin}, L. and {Chiaramida}, V. and {Chiavassa}, A. and {Chornay}, N. and {Comoretto}, G. and {Contursi}, G. and {Cooper}, W.~J. and {Cornez}, T. and {Cowell}, S. and {Crifo}, F. and {Cropper}, M. and {Crosta}, M. and {Crowley}, C. and {Dafonte}, C. and {Dapergolas}, A. and {David}, M. and {David}, P. and {de Laverny}, P. and {De Luise}, F. and {De March}, R.},
        title = "{Gaia Data Release 3. Summary of the content and survey properties}",
      journal = {\aap},
         year = 2023,
        month = jun,
       volume = {674},
          eid = {A1},
        pages = {A1},
          doi = {10.1051/0004-6361/202243940},
archivePrefix = {arXiv},
       eprint = {2208.00211},
 primaryClass = {astro-ph.GA},
       adsurl = {https://ui.adsabs.harvard.edu/abs/2023A&A...674A...1G}
}

@inproceedings{Bacon.Accardo.ea2010,
  title = {The {{MUSE}} Second-Generation {{VLT}} Instrument},
  author = {Bacon, R. and Accardo, M. and Adjali, L. and Anwand, H. and Bauer, S. and Biswas, I. and Blaizot, J. and Boudon, D. and {Brau-Nogue}, S. and Brinchmann, J. and Caillier, P. and Capoani, L. and Carollo, C. M. and Contini, T. and Couderc, P. and Daguis{\'e}, E. and Deiries, S. and Delabre, B. and Dreizler, S. and Dubois, J. and Dupieux, M. and Dupuy, C. and Emsellem, E. and Fechner, T. and Fleischmann, A. and Fran{\c c}ois, M. and Gallou, G. and Gharsa, T. and Glindemann, A. and Gojak, D. and Guiderdoni, B. and Hansali, G. and Hahn, T. and Jarno, A. and Kelz, A. and Koehler, C. and Kosmalski, J. and Laurent, F. and Le Floch, M. and Lilly, S. J. and Lizon, J. -L. and Loupias, M. and Manescau, A. and Monstein, C. and Nicklas, H. and Olaya, J. -C. and Pares, L. and Pasquini, L. and {P{\'e}contal-Rousset}, A. and Pell{\'o}, R. and Petit, C. and Popow, E. and Reiss, R. and Remillieux, A. and Renault, E. and Roth, M. and Rupprecht, G. and Serre, D. and Schaye, J. and Soucail, G. and Steinmetz, M. and Streicher, O. and Stuik, R. and Valentin, H. and Vernet, J. and Weilbacher, P. and Wisotzki, L. and Yerle, N.},
    booktitle = {Ground-based and Airborne Instrumentation for Astronomy III},
         year = 2010,
       editor = {{McLean}, Ian S. and {Ramsay}, Suzanne K. and {Takami}, Hideki},
       series = {Society of Photo-Optical Instrumentation Engineers (SPIE) Conference Series},
       volume = {7735},
        month = jul,
          eid = {773508},
        pages = {773508},
          doi = {10.1117/12.856027},
archivePrefix = {arXiv},
       eprint = {2211.16795},
 primaryClass = {astro-ph.IM},
       adsurl = {https://ui.adsabs.harvard.edu/abs/2010SPIE.7735E..08B}
}

@ARTICLE{Beers90,
       author = {{Beers}, Timothy C. and {Flynn}, Kevin and {Gebhardt}, Karl},
        title = "{Measures of Location and Scale for Velocities in Clusters of Galaxies---A Robust Approach}",
      journal = {\aj},
         year = 1990,
        month = jul,
       volume = {100},
        pages = {32},
          doi = {10.1086/115487},
       adsurl = {https://ui.adsabs.harvard.edu/abs/1990AJ....100...32B}
}

@ARTICLE{Wetzell22,
       author = {{Wetzell}, V. and {Jeltema}, T.~E. and {Hegland}, B. and {Everett}, S. and {Giles}, P.~A. and {Wilkinson}, R. and {Farahi}, A. and {Costanzi}, M. and {Hollowood}, D.~L. and {Upsdell}, E. and {Saro}, A. and {Myles}, J. and {Bermeo}, A. and {Bhargava}, S. and {Collins}, C.~A. and {Cross}, D. and {Eiger}, O. and {Gardner}, G. and {Hilton}, M. and {Jobel}, J. and {Kelly}, P. and {Laubner}, D. and {Liddle}, A.~R. and {Mann}, R.~G. and {Martinez}, V. and {Mayers}, J. and {McDaniel}, A. and {Romer}, A.~K. and {Rooney}, P. and {Sahlen}, M. and {Stott}, J. and {Swart}, A. and {Turner}, D.~J. and {Viana}, P.~T.~P. and {Abbott}, T.~M.~C. and {Aguena}, M. and {Allam}, S. and {Andrade-Oliveira}, F. and {Annis}, J. and {Asorey}, J. and {Bertin}, E. and {Burke}, D.~L. and {Calcino}, J. and {Carnero Rosell}, A. and {Carollo}, D. and {Carrasco Kind}, M. and {Carretero}, J. and {Choi}, A. and {Crocce}, M. and {da Costa}, L.~N. and {Pereira}, M.~E.~S. and {Davis}, T.~M. and {De Vicente}, J. and {Desai}, S. and {Diehl}, H.~T. and {Dietrich}, J.~P. and {Doel}, P. and {Evrard}, A.~E. and {Ferrero}, I. and {Fosalba}, P. and {Frieman}, J. and {Garc{\'\i}a-Bellido}, J. and {Gaztanaga}, E. and {Glazebrook}, K. and {Gruen}, D. and {Gruendl}, R.~A. and {Gschwend}, J. and {Gutierrez}, G. and {Hinton}, S.~R. and {Honscheid}, K. and {James}, D.~J. and {Kuehn}, K. and {Kuropatkin}, N. and {Lahav}, O. and {Lewis}, G.~F. and {Lidman}, C. and {Lima}, M. and {Maia}, M.~A.~G. and {Marshall}, J.~L. and {Melchior}, P. and {Menanteau}, F. and {Miquel}, R. and {Morgan}, R. and {Palmese}, A. and {Paz-Chinch{\'o}n}, F. and {Plazas Malag{\'o}n}, A.~A. and {Sanchez}, E. and {Scarpine}, V. and {Serrano}, S. and {Sevilla-Noarbe}, I. and {Smith}, M. and {Soares-Santos}, M. and {Suchyta}, E. and {Tarle}, G. and {Thomas}, D. and {Tucker}, B.~E. and {Tucker}, D.~L. and {Varga}, T.~N. and {Weller}, J. and {DES Collaboration}},
        title = "{Velocity dispersions of clusters in the Dark Energy Survey Y3 redMaPPer catalogue}",
      journal = {\mnras},
         year = 2022,
        month = aug,
       volume = {514},
       number = {4},
        pages = {4696-4717},
          doi = {10.1093/mnras/stac1623},
archivePrefix = {arXiv},
       eprint = {2107.07631},
 primaryClass = {astro-ph.CO},
       adsurl = {https://ui.adsabs.harvard.edu/abs/2022MNRAS.514.4696W}
}

@ARTICLE{Flowers2025,
       author = {{Flowers}, Abigail and {O'Donnell}, Jackson H. and {Jeltema}, Tesla E. and {Wetzell}, Vernon and {Roberts}, M. Grant},
        title = "{Spectroscopic and X-ray Modeling of the Strong Lensing Galaxy Cluster MACS J0138.0-2155}",
      journal = {The Open Journal of Astrophysics},
         year = 2025,
        month = jun,
       volume = {8},
          eid = {75},
        pages = {75},
          doi = {10.33232/001c.141234},
archivePrefix = {arXiv},
       eprint = {2412.19955},
 primaryClass = {astro-ph.CO},
       adsurl = {https://ui.adsabs.harvard.edu/abs/2025OJAp....8E..75F}
}

@ARTICLE{Buckly-Geer11,
       author = {{Buckley-Geer}, E.~J. and {Lin}, H. and {Drabek}, E.~R. and {Allam}, S.~S. and {Tucker}, D.~L. and {Armstrong}, R. and {Barkhouse}, W.~A. and {Bertin}, E. and {Brodwin}, M. and {Desai}, S. and {Frieman}, J.~A. and {Hansen}, S.~M. and {High}, F.~W. and {Mohr}, J.~J. and {Lin}, Y. -T. and {Ngeow}, C. -C. and {Rest}, A. and {Smith}, R.~C. and {Song}, J. and {Zenteno}, A.},
        title = "{The Serendipitous Observation of a Gravitationally Lensed Galaxy at z = 0.9057 from the Blanco Cosmology Survey: The Elliot Arc}",
      journal = {\apj},
         year = 2011,
        month = nov,
       volume = {742},
       number = {1},
          eid = {48},
        pages = {48},
          doi = {10.1088/0004-637X/742/1/48},
archivePrefix = {arXiv},
       eprint = {1108.4681},
 primaryClass = {astro-ph.CO},
       adsurl = {https://ui.adsabs.harvard.edu/abs/2011ApJ...742...48B}
}

@article{muse-pipeline,
  title = {The Data Processing Pipeline for the {{MUSE}} Instrument},
  author = {Weilbacher, Peter M. and Palsa, Ralf and Streicher, Ole and Bacon, Roland and Urrutia, Tanya and Wisotzki, Lutz and Conseil, Simon and Husemann, Bernd and Jarno, Aur{\'e}lien and Kelz, Andreas and {P{\'e}contal-Rousset}, Arlette and Richard, Johan and Roth, Martin M. and Selman, Fernando and Vernet, Jo{\"e}l},
  year = 2020,
  month = sep,
  journal = {Astronomy and Astrophysics},
  volume = {641},
  pages = {A28},
  publisher = {EDP},
  issn = {0004-6361},
  doi = {10.1051/0004-6361/202037855},
  urldate = {2025-12-09}
}

@article{scikit-learn,
  title={Scikit-learn: Machine Learning in {P}ython},
  author={Pedregosa, F. and Varoquaux, G. and Gramfort, A. and Michel, V.
          and Thirion, B. and Grisel, O. and Blondel, M. and Prettenhofer, P.
          and Weiss, R. and Dubourg, V. and Vanderplas, J. and Passos, A. and
          Cournapeau, D. and Brucher, M. and Perrot, M. and Duchesnay, E.},
  journal={Journal of Machine Learning Research},
  volume={12},
  pages={2825--2830},
  year={2011}
}

@article{Cappellari.Emsellem2004,
  title = {Parametric {{Recovery}} of {{Line-of-Sight Velocity Distributions}} from {{Absorption-Line Spectra}} of {{Galaxies}} via {{Penalized Likelihood}}},
  author = {Cappellari, Michele and Emsellem, Eric},
  year = {2004},
  month = feb,
  journal = {Publications of the Astronomical Society of the Pacific},
  volume = {116},
  pages = {138--147},
  issn = {0004-6280},
  doi = {10.1086/381875},
  urldate = {2022-12-03}
}

@article{Cappellari2017,
  title = {Improving the Full Spectrum Fitting Method: Accurate Convolution with {{Gauss-Hermite}} Functions},
  shorttitle = {Improving the Full Spectrum Fitting Method},
  author = {Cappellari, Michele},
  year = {2017},
  month = apr,
  journal = {Monthly Notices of the Royal Astronomical Society},
  volume = {466},
  pages = {798--811},
  issn = {0035-8711},
  doi = {10.1093/mnras/stw3020},
  urldate = {2022-12-03}
}

@article{Hayes.Runnholm.ea2021,
  title = {Spectral {{Shapes}} of the {{Ly$\alpha$ Emission}} from {{Galaxies}}. {{I}}. {{Blueshifted Emission}} and {{Intrinsic Invariance}} with {{Redshift}}},
  author = {Hayes, Matthew J. and Runnholm, Axel and Gronke, Max and Scarlata, Claudia},
  year = 2021,
  month = feb,
  journal = {The Astrophysical Journal},
  volume = {908},
  pages = {36},
  publisher = {IOP},
  issn = {0004-637X},
  doi = {10.3847/1538-4357/abd246},
  urldate = {2025-10-01}
}

@article{Hayes.Runnholm.ea2023,
  title = {Spectral Shapes of the {{Ly}} {$\alpha$} Emission from Galaxies - {{II}}. {{The}} Influence of Stellar Properties and Nebular Conditions on the Emergent {{Ly}} {$\alpha$} Profiles},
  author = {Hayes, Matthew J. and Runnholm, Axel and Scarlata, Claudia and Gronke, Max and {Rivera-Thorsen}, T. Emil},
  year = 2023,
  month = apr,
  journal = {Monthly Notices of the Royal Astronomical Society},
  volume = {520},
  pages = {5903--5927},
  publisher = {OUP},
  issn = {0035-8711},
  doi = {10.1093/mnras/stad477},
  urldate = {2025-10-01}
}

@article{Shapley.Steidel.ea2003,
  title = {Rest-{{Frame Ultraviolet Spectra}} of Z\textasciitilde 3 {{Lyman Break Galaxies}}},
  author = {Shapley, Alice E. and Steidel, Charles C. and Pettini, Max and Adelberger, Kurt L.},
  year = 2003,
  month = may,
  journal = {The Astrophysical Journal},
  volume = {588},
  pages = {65--89},
  publisher = {IOP},
  issn = {0004-637X},
  doi = {10.1086/373922},
  urldate = {2025-10-02}
}

@article{astropy-01,
  title = {Astropy: {{A}} Community {{Python}} Package for Astronomy},
  shorttitle = {Astropy},
  author = {{Astropy Collaboration} and Robitaille, Thomas P. and Tollerud, Erik J. and Greenfield, Perry and Droettboom, Michael and Bray, Erik and Aldcroft, Tom and Davis, Matt and Ginsburg, Adam and {Price-Whelan}, Adrian M. and Kerzendorf, Wolfgang E. and Conley, Alexander and Crighton, Neil and Barbary, Kyle and Muna, Demitri and Ferguson, Henry and Grollier, Fr{\'e}d{\'e}ric and Parikh, Madhura M. and Nair, Prasanth H. and Unther, Hans M. and Deil, Christoph and Woillez, Julien and Conseil, Simon and Kramer, Roban and Turner, James E. H. and Singer, Leo and Fox, Ryan and Weaver, Benjamin A. and Zabalza, Victor and Edwards, Zachary I. and Azalee Bostroem, K. and Burke, D. J. and Casey, Andrew R. and Crawford, Steven M. and Dencheva, Nadia and Ely, Justin and Jenness, Tim and Labrie, Kathleen and Lim, Pey Lian and Pierfederici, Francesco and Pontzen, Andrew and Ptak, Andy and Refsdal, Brian and Servillat, Mathieu and Streicher, Ole},
  year = {2013},
  month = oct,
  journal = {Astronomy and Astrophysics},
  volume = {558},
  pages = {A33},
  issn = {0004-6361},
  doi = {10.1051/0004-6361/201322068},
  urldate = {2024-08-15}
}

@article{astropy-03,
  title = {The {{Astropy Project}}: {{Sustaining}} and {{Growing}} a {{Community-oriented Open-source Project}} and the {{Latest Major Release}} (v5.0) of the {{Core Package}}},
  shorttitle = {The {{Astropy Project}}},
  author = {{Astropy Collaboration} and {Price-Whelan}, Adrian M. and Lim, Pey Lian and Earl, Nicholas and Starkman, Nathaniel and Bradley, Larry and Shupe, David L. and Patil, Aarya A. and Corrales, Lia and Brasseur, C. E. and N{\"o}the, Maximilian and Donath, Axel and Tollerud, Erik and Morris, Brett M. and Ginsburg, Adam and Vaher, Eero and Weaver, Benjamin A. and Tocknell, James and Jamieson, William and {van Kerkwijk}, Marten H. and Robitaille, Thomas P. and Merry, Bruce and Bachetti, Matteo and G{\"u}nther, H. Moritz and Aldcroft, Thomas L. and {Alvarado-Montes}, Jaime A. and Archibald, Anne M. and B{\'o}di, Attila and Bapat, Shreyas and Barentsen, Geert and Baz{\'a}n, Juanjo and Biswas, Manish and Boquien, M{\'e}d{\'e}ric and Burke, D. J. and Cara, Daria and Cara, Mihai and Conroy, Kyle E. and Conseil, Simon and Craig, Matthew W. and Cross, Robert M. and Cruz, Kelle L. and D'Eugenio, Francesco and Dencheva, Nadia and Devillepoix, Hadrien A. R. and Dietrich, J{\"o}rg P. and Eigenbrot, Arthur Davis and Erben, Thomas and Ferreira, Leonardo and {Foreman-Mackey}, Daniel and Fox, Ryan and Freij, Nabil and Garg, Suyog and Geda, Robel and Glattly, Lauren and Gondhalekar, Yash and Gordon, Karl D. and Grant, David and Greenfield, Perry and Groener, Austen M. and Guest, Steve and Gurovich, Sebastian and Handberg, Rasmus and Hart, Akeem and {Hatfield-Dodds}, Zac and Homeier, Derek and Hosseinzadeh, Griffin and Jenness, Tim and Jones, Craig K. and Joseph, Prajwel and Kalmbach, J. Bryce and Karamehmetoglu, Emir and Ka{\l}uszy{\'n}ski, Miko{\l}aj and Kelley, Michael S. P. and Kern, Nicholas and Kerzendorf, Wolfgang E. and Koch, Eric W. and Kulumani, Shankar and Lee, Antony and Ly, Chun and Ma, Zhiyuan and MacBride, Conor and Maljaars, Jakob M. and Muna, Demitri and Murphy, N. A. and Norman, Henrik and O'Steen, Richard and Oman, Kyle A. and Pacifici, Camilla and Pascual, Sergio and {Pascual-Granado}, J. and Patil, Rohit R. and Perren, Gabriel I. and Pickering, Timothy E. and Rastogi, Tanuj and Roulston, Benjamin R. and Ryan, Daniel F. and Rykoff, Eli S. and Sabater, Jose and Sakurikar, Parikshit and Salgado, Jes{\'u}s and Sanghi, Aniket and Saunders, Nicholas and Savchenko, Volodymyr and Schwardt, Ludwig and {Seifert-Eckert}, Michael and Shih, Albert Y. and Jain, Anany Shrey and Shukla, Gyanendra and Sick, Jonathan and Simpson, Chris and Singanamalla, Sudheesh and Singer, Leo P. and Singhal, Jaladh and Sinha, Manodeep and Sip{\H o}cz, Brigitta M. and Spitler, Lee R. and Stansby, David and Streicher, Ole and {\v S}umak, Jani and Swinbank, John D. and Taranu, Dan S. and Tewary, Nikita and Tremblay, Grant R. and {de Val-Borro}, Miguel and Van Kooten, Samuel J. and Vasovi{\'c}, Zlatan and Verma, Shresth and {de Miranda Cardoso}, Jos{\'e} Vin{\'i}cius and Williams, Peter K. G. and Wilson, Tom J. and Winkel, Benjamin and {Wood-Vasey}, W. M. and Xue, Rui and Yoachim, Peter and Zhang, Chen and Zonca, Andrea and {Astropy Project Contributors}},
  year = {2022},
  month = aug,
  journal = {The Astrophysical Journal},
  volume = {935},
  pages = {167},
  publisher = {IOP},
  issn = {0004-637X},
  doi = {10.3847/1538-4357/ac7c74},
  urldate = {2024-08-15}
}

@article{astropy-02,
  title = {The {{Astropy Project}}: {{Building}} an {{Open-science Project}} and {{Status}} of the v2.0 {{Core Package}}},
  shorttitle = {The {{Astropy Project}}},
  author = {{Astropy Collaboration} and {Price-Whelan}, A. M. and Sip{\H o}cz, B. M. and G{\"u}nther, H. M. and Lim, P. L. and Crawford, S. M. and Conseil, S. and Shupe, D. L. and Craig, M. W. and Dencheva, N. and Ginsburg, A. and VanderPlas, J. T. and Bradley, L. D. and {P{\'e}rez-Su{\'a}rez}, D. and {de Val-Borro}, M. and Aldcroft, T. L. and Cruz, K. L. and Robitaille, T. P. and Tollerud, E. J. and Ardelean, C. and Babej, T. and Bach, Y. P. and Bachetti, M. and Bakanov, A. V. and Bamford, S. P. and Barentsen, G. and Barmby, P. and Baumbach, A. and Berry, K. L. and Biscani, F. and Boquien, M. and Bostroem, K. A. and Bouma, L. G. and Brammer, G. B. and Bray, E. M. and Breytenbach, H. and Buddelmeijer, H. and Burke, D. J. and Calderone, G. and Cano Rodr{\'i}guez, J. L. and Cara, M. and Cardoso, J. V. M. and Cheedella, S. and Copin, Y. and Corrales, L. and Crichton, D. and D'Avella, D. and Deil, C. and Depagne, {\'E}. and Dietrich, J. P. and Donath, A. and Droettboom, M. and Earl, N. and Erben, T. and Fabbro, S. and Ferreira, L. A. and Finethy, T. and Fox, R. T. and Garrison, L. H. and Gibbons, S. L. J. and Goldstein, D. A. and Gommers, R. and Greco, J. P. and Greenfield, P. and Groener, A. M. and Grollier, F. and Hagen, A. and Hirst, P. and Homeier, D. and Horton, A. J. and Hosseinzadeh, G. and Hu, L. and Hunkeler, J. S. and Ivezi{\'c}, {\v Z}. and Jain, A. and Jenness, T. and Kanarek, G. and Kendrew, S. and Kern, N. S. and Kerzendorf, W. E. and Khvalko, A. and King, J. and Kirkby, D. and Kulkarni, A. M. and Kumar, A. and Lee, A. and Lenz, D. and Littlefair, S. P. and Ma, Z. and Macleod, D. M. and Mastropietro, M. and McCully, C. and Montagnac, S. and Morris, B. M. and Mueller, M. and Mumford, S. J. and Muna, D. and Murphy, N. A. and Nelson, S. and Nguyen, G. H. and Ninan, J. P. and N{\"o}the, M. and Ogaz, S. and Oh, S. and Parejko, J. K. and Parley, N. and Pascual, S. and Patil, R. and Patil, A. A. and Plunkett, A. L. and Prochaska, J. X. and Rastogi, T. and Reddy Janga, V. and Sabater, J. and Sakurikar, P. and Seifert, M. and Sherbert, L. E. and {Sherwood-Taylor}, H. and Shih, A. Y. and Sick, J. and Silbiger, M. T. and Singanamalla, S. and Singer, L. P. and Sladen, P. H. and Sooley, K. A. and Sornarajah, S. and Streicher, O. and Teuben, P. and Thomas, S. W. and Tremblay, G. R. and Turner, J. E. H. and Terr{\'o}n, V. and {van Kerkwijk}, M. H. and {de la Vega}, A. and Watkins, L. L. and Weaver, B. A. and Whitmore, J. B. and Woillez, J. and Zabalza, V. and {Astropy Contributors}},
  year = {2018},
  month = sep,
  journal = {The Astronomical Journal},
  volume = {156},
  pages = {123},
  publisher = {IOP},
  issn = {0004-6256},
  doi = {10.3847/1538-3881/aabc4f},
  urldate = {2024-08-15}
}

@article{numpy,
  title = {Array Programming with {{NumPy}}},
  author = {Harris, Charles R. and Millman, K. Jarrod and {van der Walt}, St{\'e}fan J. and Gommers, Ralf and Virtanen, Pauli and Cournapeau, David and Wieser, Eric and Taylor, Julian and Berg, Sebastian and Smith, Nathaniel J. and Kern, Robert and Picus, Matti and Hoyer, Stephan and {van Kerkwijk}, Marten H. and Brett, Matthew and Haldane, Allan and {del R{\'i}o}, Jaime Fern{\'a}ndez and Wiebe, Mark and Peterson, Pearu and {G{\'e}rard-Marchant}, Pierre and Sheppard, Kevin and Reddy, Tyler and Weckesser, Warren and Abbasi, Hameer and Gohlke, Christoph and Oliphant, Travis E.},
  year = {2020},
  month = sep,
  journal = {Nature},
  volume = {585},
  pages = {357--362},
  issn = {0028-0836},
  doi = {10.1038/s41586-020-2649-2},
  urldate = {2024-08-15}
}

@article{matplotlib,
  title = {Matplotlib: {{A 2D Graphics Environment}}},
  shorttitle = {Matplotlib},
  author = {Hunter, John D.},
  year = {2007},
  month = may,
  journal = {Computing in Science and Engineering},
  volume = {9},
  pages = {90--95},
  doi = {10.1109/MCSE.2007.55},
  urldate = {2024-08-15}
}

@article{scipy,
  title = {{{SciPy}} 1.0: Fundamental Algorithms for Scientific Computing in {{Python}}},
  shorttitle = {{{SciPy}} 1.0},
  author = {Virtanen, Pauli and Gommers, Ralf and Oliphant, Travis E. and Haberland, Matt and Reddy, Tyler and Cournapeau, David and Burovski, Evgeni and Peterson, Pearu and Weckesser, Warren and Bright, Jonathan and {van der Walt}, St{\'e}fan J. and Brett, Matthew and Wilson, Joshua and Millman, K. Jarrod and Mayorov, Nikolay and Nelson, Andrew R. J. and Jones, Eric and Kern, Robert and Larson, Eric and Carey, C. J. and Polat, {\.I}lhan and Feng, Yu and Moore, Eric W. and VanderPlas, Jake and Laxalde, Denis and Perktold, Josef and Cimrman, Robert and Henriksen, Ian and Quintero, E. A. and Harris, Charles R. and Archibald, Anne M. and Ribeiro, Ant{\^o}nio H. and Pedregosa, Fabian and {van Mulbregt}, Paul and {SciPy 1. 0 Contributors}},
  year = {2020},
  month = feb,
  journal = {Nature Methods},
  volume = {17},
  pages = {261--272},
  doi = {10.1038/s41592-019-0686-2},
  urldate = {2024-08-15}
}

@article{Manjon-Garcia.Diego.ea2020a,
  title = {Constraining the Abundance of Dark Matter in the Central Region of the Galaxy Cluster {{MACS J1206}}.2-0847 with a Free-Form Strong Lensing Analysis},
  author = {{Manj{\'o}n-Garc{\'i}a}, Alberto and Diego, Jose M. and Herranz, Diego and Lam, Daniel},
  year = 2020,
  month = jul,
  journal = {Astronomy and Astrophysics},
  volume = {639},
  pages = {A125},
  publisher = {EDP},
  issn = {0004-6361},
  doi = {10.1051/0004-6361/201936914},
  urldate = {2026-01-12}
}

@misc{Cerny.Jauzac.ea2025,
  title = {The {{Kaleidoscope Survey}}: {{Strong Gravitational Lensing}} in {{Galaxy Clusters}} with {{Radial Arcs}}},
  shorttitle = {The {{Kaleidoscope Survey}}},
  author = {Cerny, Catherine and Jauzac, Mathilde and Lagattuta, David and Niemiec, Anna and Mahler, Guillaume and Edge, Alastair and Massey, Richard},
  year = 2025,
  month = jun,
  number = {arXiv:2506.21531},
  eprint = {2506.21531},
  primaryclass = {astro-ph},
  publisher = {arXiv},
  doi = {10.48550/arXiv.2506.21531},
  urldate = {2025-06-27},
  archiveprefix = {arXiv}
}

@article{Bulbul.Liu.ea2024,
  title = {The {{SRG}}/{{eROSITA All-Sky Survey}}. {{The}} First Catalog of Galaxy Clusters and Groups in the {{Western Galactic Hemisphere}}},
  author = {Bulbul, E. and Liu, A. and Kluge, M. and Zhang, X. and Sanders, J. S. and Bahar, Y. E. and Ghirardini, V. and Artis, E. and Seppi, R. and Garrel, C. and {Ramos-Ceja}, M. E. and Comparat, J. and Balzer, F. and B{\"o}ckmann, K. and Br{\"u}ggen, M. and Clerc, N. and Dennerl, K. and Dolag, K. and Freyberg, M. and Grandis, S. and Gruen, D. and Kleinebreil, F. and Krippendorf, S. and Lamer, G. and Merloni, A. and Migkas, K. and Nandra, K. and Pacaud, F. and Predehl, P. and Reiprich, T. H. and Schrabback, T. and Veronica, A. and Weller, J. and Zelmer, S.},
  year = 2024,
  month = may,
  journal = {Astronomy and Astrophysics},
  volume = {685},
  pages = {A106},
  publisher = {EDP},
  issn = {0004-6361},
  doi = {10.1051/0004-6361/202348264},
  urldate = {2026-01-14}
}

@article{Klein.Mohr.ea2024,
  title = {The {{ACT-DR5 MCMF}} Galaxy Cluster Catalog},
  author = {Klein, M. and Mohr, J. J. and Davies, C. T.},
  year = 2024,
  month = oct,
  journal = {Astronomy and Astrophysics},
  volume = {690},
  pages = {A322},
  publisher = {EDP},
  issn = {0004-6361},
  doi = {10.1051/0004-6361/202451203},
  urldate = {2026-01-13}
}

@article{Ludlow.Bose.ea2016a,
  title = {The Mass-Concentration-Redshift Relation of Cold and Warm Dark Matter Haloes},
  author = {Ludlow, Aaron D. and Bose, Sownak and Angulo, Ra{\'u}l E. and Wang, Lan and Hellwing, Wojciech A. and Navarro, Julio F. and Cole, Shaun and Frenk, Carlos S.},
  year = {2016},
  month = aug,
  journal = {Monthly Notices of the Royal Astronomical Society},
  volume = {460},
  pages = {1214--1232},
  publisher = {OUP},
  issn = {0035-8711},
  doi = {10.1093/mnras/stw1046},
  urldate = {2025-07-27}
}

@article{Gilmore.Natarajan2009,
  title = {Cosmography with Cluster Strong Lensing},
  author = {Gilmore, James and Natarajan, Priyamvada},
  year = 2009,
  month = jun,
  journal = {Monthly Notices of the Royal Astronomical Society},
  volume = {396},
  pages = {354--364},
  publisher = {OUP},
  issn = {0035-8711},
  doi = {10.1111/j.1365-2966.2009.14612.x},
  urldate = {2026-01-20}
}

@article{McLinden.Finkelstein.ea2011,
  title = {First {{Spectroscopic Measurements}} of [{{O III}}] {{Emission}} from {{Ly$\alpha$ Selected Field Galaxies}} at z \textasciitilde{} 3.1},
  author = {McLinden, Emily M. and Finkelstein, Steven L. and Rhoads, James E. and Malhotra, Sangeeta and Hibon, Pascale and Richardson, Mark L. A. and Cresci, Giovanni and Quirrenbach, Andreas and Pasquali, Anna and Bian, Fuyan and Fan, Xiaohui and Woodward, Charles E.},
  year = 2011,
  month = apr,
  journal = {The Astrophysical Journal},
  volume = {730},
  pages = {136},
  publisher = {IOP},
  issn = {0004-637X},
  doi = {10.1088/0004-637X/730/2/136},
  urldate = {2026-01-20}
}

@misc{PaperII,
  title = {The {{Carousel Lens II}}: {{Cosmological Constraints}} with {{GIGA-Lens}}},
  shorttitle = {The {{Carousel Lens II}}},
  author = {Urcelay, Felipe and Huang, Xiaosheng and Sheu, William and O'Donnell, Jackson H. and Jeltema, Tesla and Williams, Demetrius Y. and Xu, Sean and Agarwal, Shrihan and Aldering, Greg and {\'A}lvarez-Garc{\'i}a, David and Ambardekar, Harsh and Barone, Tania M. and Bian, Fuyan and Bolton, Adam S. and Cikota, Aleksandar and Farren, Gerrit S. and Glazebrook, Karl and Hoyt, Taylor and Jain, Aniket and Jones, Tucker and Kacprzak, Glenn G. and Lin, Emerald and Perlmutter, Saul and Rubin, David and Schlegel, David J. and Silver, Ethan and Storfer, Christopher J. and Suzuki, Nao and Truong, Jannik and {\'U}beda, M{\'o}nica and C, Keerthi Vasan G.},
  year = 2026,
  month = feb,
  publisher = {arXiv},
  doi = {10.48550/arXiv.2602.16077},
  urldate = {2026-03-04}
}

@ARTICLE{Anbajagane22,
       author = {{Anbajagane}, Dhayaa and {Aung}, Han and {Evrard}, August E. and {Farahi}, Arya and {Nagai}, Daisuke and {Barnes}, David J. and {Cui}, Weiguang and {Dolag}, Klaus and {McCarthy}, Ian G. and {Rasia}, Elena and {Yepes}, Gustavo},
        title = "{Galaxy velocity bias in cosmological simulations: towards per cent-level calibration}",
      journal = {\mnras},
         year = 2022,
        month = feb,
       volume = {510},
       number = {2},
        pages = {2980-2997},
          doi = {10.1093/mnras/stab3587},
archivePrefix = {arXiv},
       eprint = {2110.01683},
 primaryClass = {astro-ph.CO},
       adsurl = {https://ui.adsabs.harvard.edu/abs/2022MNRAS.510.2980A}
}

\end{document}